\documentclass[journal]{IEEEtran}

\usepackage{graphicx}
\usepackage{amsmath,bbm,epsfig,amssymb,amsfonts,  amstext, verbatim,amsopn,cite,subfigure,multirow,multicol,lipsum}
\usepackage{balance}
\usepackage{url}
\usepackage{amsfonts}
\usepackage{epsfig}
\usepackage{epstopdf}
\usepackage{setspace}
\usepackage{psfrag}	
\usepackage{multirow}
\usepackage{float}
\usepackage[process=auto]{pstool}
\usepackage{etoolbox}
\usepackage{algorithm}
\usepackage{algorithmic}
\allowdisplaybreaks

\usepackage[T1]{fontenc} 
\usepackage{amsmath}
\usepackage{bm} 

\newcommand*{\QED}{\hfill\ensuremath{\square}}%

\newtheorem{lem}{Lemma}
\newtheorem{remk}{Remark}

\newtheorem{corol}{Corollary}

\newtoggle{OneColumn}

\togglefalse{OneColumn}

\iftoggle{OneColumn}{%
  
}{%
}

\EndPreamble
\begin{document}

\title{Channel Estimation for Diffusive Molecular Communications }

\author{Vahid Jamali, \textit{Student Member, IEEE}, Arman Ahmadzadeh, \textit{Student Member, IEEE}, Christophe Jardin, \\ Heinrich Sticht, and Robert Schober, \textit{Fellow, IEEE} \vspace{-0.1cm}\\ 
\thanks{ This paper was presented in part at the International Conference on Communications (ICC) 2016, Kuala Lumpur, Malaysia [31].}
\thanks{V. Jamali, A. Ahmadzadeh, and R. Schober are with the Institute for Digital Communications
at  the  Friedrich-Alexander  University  (FAU),  Erlangen,  Germany  (email:
vahid.jamali@fau.de, arman.ahmadzadeh@fau.de, and robert.schober@fau.de).}  
\thanks{C. Jardin and H. Sticht are with the Institute for Biochemistry
at  the  Friedrich-Alexander  University  (FAU),  Erlangen,  Germany  (email: christophe.jardin@fau.de
and heinrich.sticht@fau.de).} 
}

\maketitle

\begin{abstract}

In molecular communication (MC) systems, the \textit{expected} number of molecules observed at the receiver over time after the instantaneous release of molecules by the transmitter is referred to as the channel impulse response (CIR). Knowledge of the CIR is needed for the design of detection and equalization schemes. In this paper, we  present a training-based CIR estimation framework for MC systems which aims at estimating the CIR based on the \textit{observed} number of molecules at the receiver due to emission of a \textit{sequence} of known numbers of molecules by the transmitter. Thereby, we distinguish two scenarios depending on whether or not statistical channel knowledge is available.  In particular,  we derive maximum likelihood (ML) and least sum of square errors (LSSE)  estimators which do not require any knowledge of the channel statistics. For the case, when statistical channel knowledge is available, the corresponding maximum a posteriori (MAP) and linear minimum mean square error (LMMSE) estimators are provided. As performance bound, we derive the classical Cramer Rao (CR) lower bound, valid for any unbiased estimator, which does not exploit statistical channel knowledge, and the Bayesian CR lower bound, valid for any unbiased estimator, which exploits statistical channel knowledge.  Finally, we propose optimal and suboptimal  training sequence designs for the considered MC system.  Simulation results confirm the analysis and compare the performance of the proposed estimation techniques with the respective CR lower bounds. 

\end{abstract}


\section{Introduction}

Recent advances in biology, nanotechnology, and medicine have enabled the possibility of communication in nano/micrometer scale environments \cite{Survey_Mol_Net,Nariman_Survey,Survey_Mol_Nono}.  Thereby,   employing molecules as information carriers, molecular communication (MC) has quickly emerged as a bio-inspired approach for man-made communication systems in such  environments. In fact, calcium signaling among neighboring cells, the use of neurotransmitters for communication across the synaptic cleft of neurons, and the exchange of autoinducers as signaling molecules in bacteria for quorum sensing are among the many examples of MC in nature \cite{CellBio,Survey_Mol_Nono}.

\subsection{Motivation} 

 The design of any communication system crucially depends on the characteristics of the channel under consideration. In MC systems, the impact of the channel on
the number of observed molecules can be captured by the channel impulse response (CIR) which is defined as the \textit{expected} number of  molecules counted at the receiver at time $t$ after the instantaneous release of a known number of molecules by the transmitter at time $t=0$. The CIR, denoted by $\bar{c}(t)$, can be used as the basis for the design of   equalization and detection schemes for MC systems \cite{Adam_OptReciever,ConsCIR}. For diffusion-based MC, the  released molecules move randomly according to Brownian motion  which is caused by thermal vibration and collisions with other  molecules in the fluid environment. Thereby, the average concentration of the molecules at a given coordinate $\mathbf{a}=[a_x, a_y, a_z]$ and at time $t$ after their release by the transmitter, denoted by $\bar{\mathcal{C}}(\mathbf{a},t)$, is governed by Fick's second law of diffusion \cite{Adam_OptReciever}.  Finding $\bar{\mathcal{C}}(\mathbf{a},t)$ analytically involves solving partial differential equations and depends on initial and boundary conditions. Therefore, one possible approach for determining the CIR, which is widely employed in the literature \cite{ConsCIR}, is to first derive a sufficiently accurate analytical expression for $\bar{\mathcal{C}}(\mathbf{a},t)$ for the considered MC channel from Fick's second law, and to subsequently integrate it over the receiver volume, $V^{\mathtt{rec}}$, i.e., 
\begin{IEEEeqnarray}{lll} \label{Eq:Cons_CIR}
  \bar{c}(t) = \iiint_{\mathbf{a}\in V^{\mathtt{rec}}} \bar{\mathcal{C}}(\mathbf{a},t) \mathrm{d}a_x\mathrm{d}a_y\mathrm{d}a_z.
\end{IEEEeqnarray}
However, this approach may not be applicable in many practical scenarios as discussed in the following.

\begin{itemize}
\item The CIR can be obtained based on (\ref{Eq:Cons_CIR}) only for the special case of a fully \textit{transparent} receiver where it is assumed that the molecules move through the receiver as if it was not present in the environment. The assumption of a fully transparent receiver  is a valid approximation only for some particular scenarios where the interaction of the receiver with the molecules can be neglected. However, for general receivers, the relationship between the concentration $\bar{\mathcal{C}}(\mathbf{a},t)$  and the number of observed  molecules $\bar{c}(t)$ may not be as straightforward \cite{Chae_Absorbing,Arman_ReactReciever,Yilmaz_Absorb}. 
\item Solving the differential equation associated with Fick's second law is possible only for simple and idealistic environments.  For example, assuming a \textit{point} source located at the origin of an \textit{unbounded} environment and \textit{impulsive} molecule release, $\bar{\mathcal{C}}(\mathbf{a},t)$ is obtained as \cite{ConsCIR}
\begin{IEEEeqnarray}{lll} \label{Eq:Consentration}
  \bar{\mathcal{C}}(\mathbf{a},t) = \frac{N^{\mathtt{Tx}}}{\left(4\pi D t\right)^{3/2}} \exp\left(-\frac{|\mathbf{a}|^2}{4Dt}\right)\quad \left[\frac{\text{\small molecules}}{\text{\small m}^3}\right], \,\,\,
\end{IEEEeqnarray}
where $N^{\mathtt{Tx}}$ is the number of molecules released  by the transmitter at $t=0$ and $D$ is the diffusion coefficient of the signaling molecule.  However, $\bar{\mathcal{C}}(\mathbf{a},t)$ cannot be obtained in closed form for most practical MC environments which may involve difficult boundary conditions, non-instantaneous molecule release, flow, etc.  Additionally, as has been shown in  \cite{Wil_Nature},  the classical Fick's diffusion equation  might even not be  applicable in  complex MC environments as  physico-chemical interactions of the molecules with other objects in the channel, such as other molecules, cells, and microvessels, are not accounted for.

\item Even if an expression for $\bar{\mathcal{C}}(\mathbf{a},t)$ can be obtained for a particular MC system, e.g. (\ref{Eq:Consentration}), it will be a function of several channel parameters such as the distance between the transmitter and the receiver and the diffusion coefficient. However, in practice, these parameters may  not be known a priori and also have to be estimated \cite{MC_Distance,AdamDistanceEstimation}. This complicates finding the CIR based on $\bar{\mathcal{C}}(\mathbf{a},t)$. 

\end{itemize}

Fortunately, for most communication problems, including equalization and detection, only the \textit{expected} number of molecules that the receiver observes at the sampling times is needed \cite{Adam_OptReciever,ConsCIR}. Hence, the difficulties associated with deriving $\bar{\mathcal{C}}(\mathbf{a},t)$ can be avoided by directly estimating the CIR. 
 Motivated by the above discussion, our goal in this paper is to develop a general CIR estimation framework for  MC systems which is not limited to a particular MC channel model or a specific receiver type  and does not  require knowledge of the  channel parameters.

 
\subsection{Related Work} 

In most existing works on MC, the CIR is assumed to be perfectly known for  receiver design \cite{Adam_OptReciever,ConsCIR,Arman_AF}. In the following, we  review the relevant MC literature that focused on channel characterization. Estimation of the distance between a transmitter and a receiver  was studied in \cite{MC_Distance,DistanceEstLett,AdamDistanceEstimation} for diffusive MC. In \cite{Akyildiz_MC_E2E}, an end-to-end mathematical  model, including  transmitter,  channel, and  receiver, was presented, and in \cite{MC_Stoch_Model}, a stochastic channel model was proposed for  flow-based and diffusion-based MC. For active transport MC, a Markov chain channel model was derived in \cite{Farsad_Markov}. Additionally, a unifying model including the effects of external noise sources and inter-symbol interference (ISI) was proposed for diffusive MC  in \cite{Adam_Universal_Noise}.  In  \cite{Akyildiz_MC_Memory}, the authors analyzed a microfluidic MC channel, propagation noise, and channel memory. For neuro-spike communications, a physical model for the channel between two neurons was developed in \cite{Chn_Neuro}. In addition to these theoretical works, experimental results for the characterization of MC channels were reported in \cite{MC_test_Eckford,MC_test_Zanella}.  However, the focus  of \cite{MC_Distance,AdamDistanceEstimation,Akyildiz_MC_E2E,MC_Stoch_Model,Farsad_Markov,Adam_Universal_Noise,Akyildiz_MC_Memory,Chn_Neuro,MC_test_Eckford,MC_test_Zanella} is either channel modeling or the estimation of channel parameters, i.e., the obtained results are not directly applicable to CIR acquisition. In contrast to MC, for conventional wireless communication, there is a rich literature on channel estimation, mainly for linear channel models and impairment by additive white Gaussian noise (AWGN), see \cite{LSSE,Channel_Estimation_Arsalan,Channel_Estimation_Li}, and the references therein.  Channel estimation was also studied for non-linear and/or non-AWGN channels especially  in  optical communication. For example, a linear time-invariant system with chi-squared noise was used in \cite{Optic_Chi_Channel} to model the optical fiber channel. However, the MC channel model considered in this paper is different from the channel models considered in \cite{LSSE,Channel_Estimation_Arsalan,Channel_Estimation_Li,Optic_Chi_Channel}, and hence, these results are not directly applicable to MC.

In this paper, we consider the Poisson channel which was introduced in \cite{Hamid_Lett,HamidJSAC} for molecule-counting receivers in MC systems.  The Poisson channel was previously used to model optical communication channels with photon-counting receivers, see \cite{Poisson_Channel} and the references therein. Moreover, the authors of a recent paper \cite{Channel_Estimation_Optic_ISI} adopted the Poisson channel model for non-line-of-sight (NLOS) optical wireless communication and proposed two different channel estimators. However, to the best of the authors' knowledge, channel estimation for the Poisson MC channel was first considered in the conference version of this paper \cite{ICC2016_MC_Arxiv}.  We note that the channel estimators developed in this paper for MC systems are different from the channel estimators presented in \cite{Channel_Estimation_Optic_ISI} for NLOS optical communications. Moreover, the estimators considered in \cite{Channel_Estimation_Optic_ISI} do not take into account the  non-negativity of the CIR coefficients. In contrast, we have carefully considered this non-negativity constraint on the CIR coefficients in the proposed estimators. Finally, in contrast to \cite{Channel_Estimation_Optic_ISI}, we assume that the mean of the noise is unknown and has to be estimated as well.  Therefore, the results in \cite{Channel_Estimation_Optic_ISI} are not directly applicable to the MC channel considered in this paper.

\begin{figure*}[!ht]
  \centering
    \begin{minipage}[b]{0.45\textwidth}
  \centering
\raisebox{0.6\height}{ \scalebox{0.8}{
\pstool[width=1\linewidth]{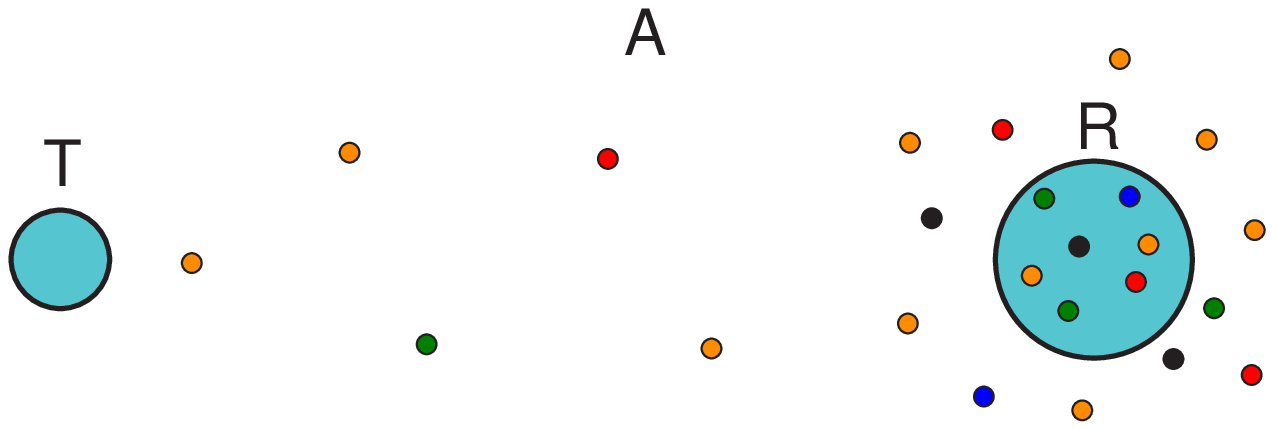}{
\psfrag{T}[c][c][1.2]{$\mathtt{Tx}$}
\psfrag{R}[c][c][1.2]{$\mathtt{Rx}$}
\psfrag{A}[l][c][1]{a)}
} } }
  \end{minipage}
    \hfill
    \begin{minipage}[b]{0.5\textwidth}
  \centering
\resizebox{0.8\linewidth}{!}{\psfragfig{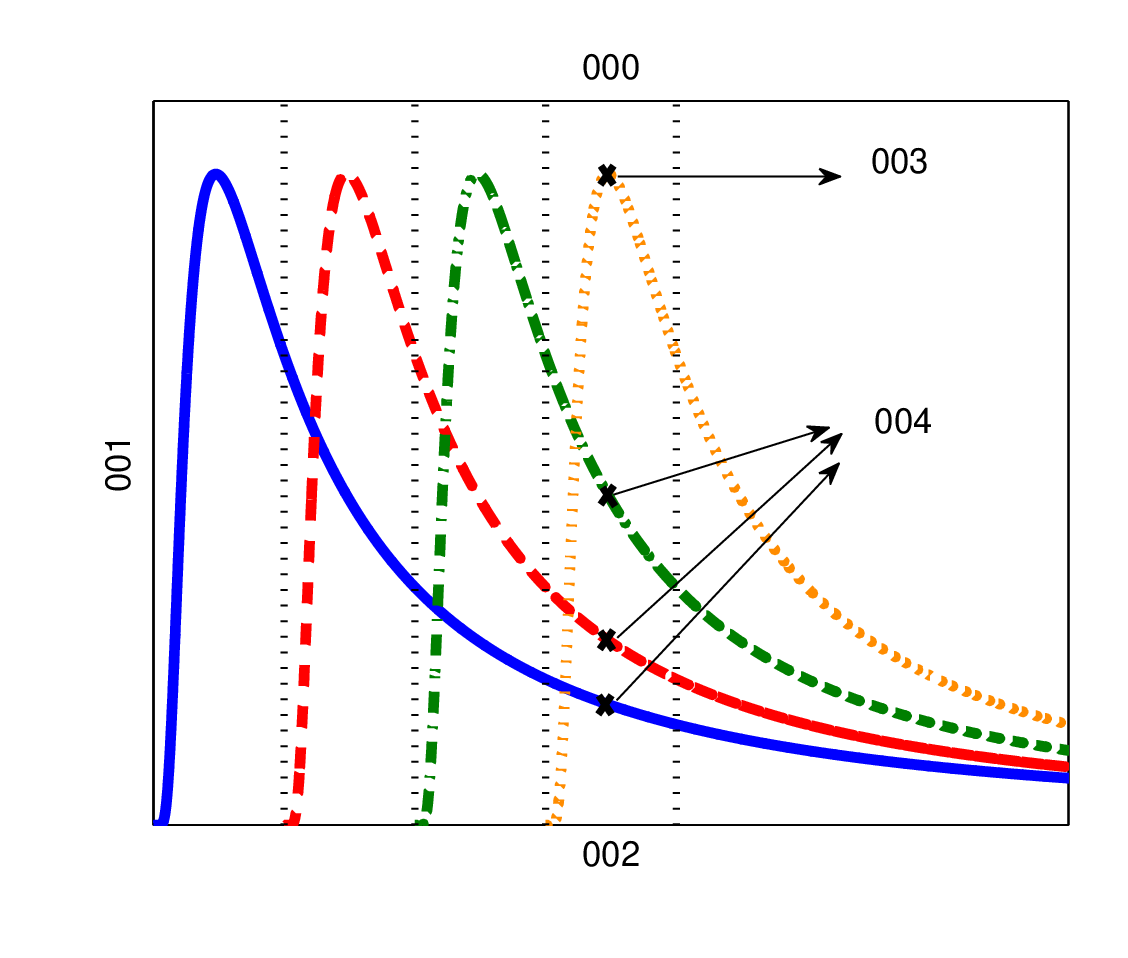}} 
  \end{minipage}
\caption{a) An illustrative time snapshot of an MC system during the fourth symbol interval where the transmitted molecules in the first, second, third, and fourth symbol intervals are shown in  blue, red, green, and orange  colors, respectively, and  noise molecules are shown in black color. b) Illustration of ISI: the expected number of observed molecules at the receiver if the transmitter emits $N^{\mathtt{Tx}}$ molecules at the beginning of each symbol interval. The vertical dotted lines indicate the beginning of a new symbol interval.   }
\label{Fig:ISI}
\end{figure*}

\subsection{Contributions} 
 
In this paper, we directly estimate the CIR of the MC channel based on the channel output, i.e., the number of molecules observed at the receiver.  In particular, we present a training-based CIR estimation framework which aims at estimating the CIR based on the detected number of molecules at the receiver due to the emission of a  sequence of known numbers of molecules by the transmitter. Thereby, we consider the following two scenarios: \textit{i)} Non-Bayesian CIR estimation where statistical CIR knowledge is not available or not exploited. For this case, we first derive the optimal maximum likelihood (ML) CIR  estimator. Subsequently, we obtain the suboptimal least sum of square errors (LSSE) CIR  estimator which entails a lower computational complexity than the ML estimator. Additionally, we derive the classical Cramer Rao (CR) bound which constitutes a lower bound on the estimation error variance of any unbiased estimator. \textit{ii)} Bayesian CIR estimation where knowledge of the CIR statistics  is available and exploited. Here, we first investigate   maximum a posteriori (MAP)  estimation as the optimal approach which requires full statistical knowledge of the MC channel. Subsequently, we derive the linear minimum mean square error (LMMSE)  estimator for the case when only the first and second order statistics of the MC channel are available. Additionally, we derive the Bayesian CR bound as a generalization of the classical CR bound for estimators which exploit statistical channel knowledge. Finally, we study optimal and suboptimal  training sequence designs for the considered MC system. Simulation results  confirm the analysis and evaluate the performance of the proposed estimation techniques with respect to  the corresponding CR lower bound.

We note that in contrast to the conference version \cite{ICC2016_MC_Arxiv}, which studies only the case when statistical channel knowledge is not exploited, this paper also provides optimal and suboptimal CIR estimators, training sequence designs, and performance bounds for the case when statistical channel knowledge is exploited. Moreover, many of the extensive discussions, simulation results, and proofs provided in this paper are not included in \cite{ICC2016_MC_Arxiv}.

\subsection{Organization and Notation}

The remainder of this paper is organized as follows. In Section~II, some preliminaries and assumptions are presented and the classical and Bayesian CR bounds are derived. CIR estimators which do not require statistical channel knowledge are proposed in Section~III, and CIR estimators which do take advantage of statistical channel knowledge are given in Section~IV. In Section~V, several different training sequence designs are presented. Numerical results are provided in Section~VI, and conclusions are drawn in Section~VII.
  
\textit{Notations:} We use the following notations throughout this paper: $\mathbbmss{E}_{x}\{\cdot\}$ denotes expectation with respect to random variable (RV) $x$, $|\cdot|$ represents the cardinality of a set,  and $[x]^+=\max\{0,x\}$.  Bold capital and small letters are used to denote matrices and vectors, respectively. $\mathbf{1}$ and $\mathbf{0}$ are vectors whose elements are all ones and zeros, respectively,  $\mathbf{I}$ denotes the identity matrix, $\mathbf{A}^T$ denotes the transpose of $\mathbf{A}$,  $\|\mathbf{a}\|$ represents the norm of  vector $\mathbf{a}$, $[\mathbf{A}]_{mn}$ denotes the element in the $m$-th row and $n$-th column of matrix $\mathbf{A}$, $\mathrm{tr}\{\mathbf{A}\}$ is the trace of matrix $\mathbf{A}$, $\mathrm{diag}\{\mathbf{a}\}$ denotes a diagonal matrix with the elements of vector $\mathbf{a}$ on its main diagonal, $\mathrm{vdiag}\{\mathbf{A}\}$ is a vector which contains the diagonal entries
of  matrix $\mathbf{A}$, $\mathrm{eig}\{\mathbf{A}\}$ is the set of eigen-values of matrix $\mathbf{A}$,  $\mathbf{A}\succeq 0$  denotes a positive semidefinite matrix $\mathbf{A}$, and $\mathbf{a}\geq \mathbf{0}$ specifies that all elements of vector $\mathbf{a}$ are non-negative. Additionally, $\nabla^2_{\mathbf{a}\mathbf{a}} f(\mathbf{a})$ denotes the Hessian matrix of function $f(\mathbf{a})$ with respect to vector $\mathbf{a}$. Furthermore,  $\mathrm{Poiss}(\lambda)$  denotes a Poisson RV with mean  $\lambda$, $\mathrm{Bin}(n,p)$ denotes a binomial RV for $n$  trials and success probability $p$, and $\mathcal{N}\left(\mathbf{a},\mathbf{A}\right)$ denotes a multivariate normal RV with mean vector $\mathbf{a}$ and covariance matrix~$\mathbf{A}$.

\section{Preliminaries, Assumptions, and Performance Bounds}

In this section, we present the considered MC channel model,  formulate the CIR estimation problem, and derive the corresponding classical and Bayesian CR bounds.  

\subsection{System Model}

We consider an MC system  consisting of a transmitter, a channel, and a receiver, see Fig.~\ref{Fig:ISI}~a). At the beginning of each symbol interval, the transmitter releases a fraction of $N^{\mathtt{Tx}}$ molecules where $N^{\mathtt{Tx}}$ is the maximum number of molecules that the transmitter can release at once, i.e., concentration shift keying (CSK) is performed \cite{Nariman_Survey}. In this paper, we assume that the transmitter emits only one type of molecule. The released molecules propagate through the medium between the transmitter and the receiver. We assume that the movements of individual molecules are independent from each other. The receiver counts the number of observed molecules in each symbol interval.  We note that this is a rather general model for the MC receiver which includes well-known receivers such as the transparent receiver \cite{ConsCIR}, the absorbing receiver \cite{Chae_Absorbing}, and the reactive receiver \cite{Arman_ReactReciever}.

Due to the memory of the MC channel, inter-symbol interference (ISI) occurs \cite{Adam_Universal_Noise,Akyildiz_MC_Memory}, see Fig.~\ref{Fig:ISI}~b). In particular, ISI-free communication is only possible  if  the symbol intervals are chosen sufficiently large such that the CIR fully decays to zero within one symbol interval which severely limits the transmission rate and results in an inefficient MC system design. Therefore, taking into account the effect of ISI, we assume the following input-output relation for the MC system
\begin{IEEEeqnarray}{lll} \label{Eq:ChannelInOut}
  r[k]  = \sum_{l=1}^{L} c_l[k] + c_{\mathtt{n}}[k],
\end{IEEEeqnarray}
where $r[k]$ is the number of molecules detected at the receiver in symbol interval $k$, $L$ is the number of memory taps of the channel, and $c_l[k]$ is the number of  molecules observed at the receiver in symbol interval $k$ due to the release of $s[k-l+1]N^{\mathtt{Tx}}$ molecules by the transmitter in symbol interval $k-l+1$, where $s[k]\in[0,1]$ is the transmitted symbol in symbol interval~$k$.
Thereby,  $c_l[k]$ can be well approximated by a Poisson RV with mean $\bar{c}_l s[k-l+1]$, i.e., $c_l[k]\sim\mathrm{Poiss}\left(\bar{c}_l s[k-l+1]\right)$, see \cite{Hamid_Lett,Adam_OptReciever}. Moreover, $c_{\mathtt{n}}[k]$ is the number of external noise molecules detected by the receiver in symbol interval $k$  but not released by the transmitter. Noise molecules may originate from  interfering sources which employ the same type of molecule as the considered MC system. Hence, $c_{\mathtt{n}}[k]$ can  also be modeled as a Poisson  RV, i.e., $c_{\mathtt{n}}[k]\sim\mathrm{Poiss}\left(\bar{c}_{\mathtt{n}}\right)$, where $\bar{c}_{\mathtt{n}}=\mathbbmss{E}\left\{c_{\mathtt{n}}[k]\right\}$. 

\begin{remk}
From a probabilistic point of view, we can assume that each molecule released by the transmitter in symbol interval $k-l+1$ is observed at the receiver in symbol interval $k$ with a certain probability, denoted by $p_l$. Thereby, the probability that $n$ molecules are observed at the receiver  in symbol interval $k$ due to the emission of $N^{\mathtt{Tx}}$ molecules in symbol interval $k-l+1$ follows a binomial distribution, i.e., $n\sim\mathrm{Bin}(N^{\mathtt{Tx}},p_l)$. Moreover, assuming $N^{\mathtt{Tx}}\to\infty$ while $N^{\mathtt{Tx}}p_l\triangleq \bar{c}_l$ is fixed, the  binomial distribution $\mathrm{Bin}(N^{\mathtt{Tx}},p_l)$ converges to the Poisson distribution $\mathrm{Poiss}(\bar{c}_l)$ \cite{BayesianBook}. This  is a  valid assumption in MC since the number of released molecules is often very large to ensure that a sufficient number of molecules reaches the receiver. The same reasoning applies to the noise molecules. \hfill $\QED$
\end{remk}

\begin{figure}
\centering
\scalebox{0.55}{
\pstool[width=1.8\linewidth]{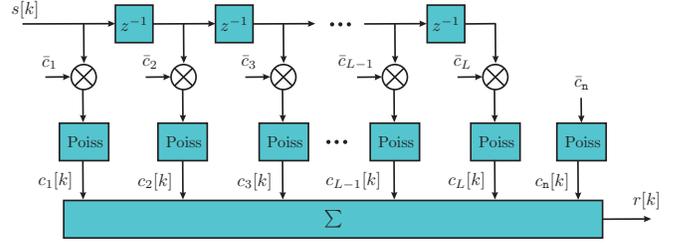}{
\psfrag{Z}[c][c][1.2]{$z^{-1}$}
\psfrag{C0}[c][c][1.2]{$\bar{c}_1$}
\psfrag{C1}[c][c][1.2]{$\bar{c}_2$}
\psfrag{C2}[c][c][1.2]{$\bar{c}_3$}
\psfrag{Cl}[c][c][1.2]{$\bar{c}_{L-1}$}
\psfrag{CL}[c][c][1.2]{$\bar{c}_L$}
\psfrag{Cn}[c][c][1.2]{$\bar{c}_{\mathtt{n}}$}
\psfrag{u0}[c][c][1.2]{$c_1[k]$}
\psfrag{u1}[c][c][1.2]{$c_2[k]$}
\psfrag{u2}[c][c][1.2]{$c_3[k]$}
\psfrag{ul}[c][c][1.2]{$c_{L-1}[k]$}
\psfrag{uL}[c][c][1.2]{$c_L[k]$}
\psfrag{st}[c][c][1.2]{$s[k]$}
\psfrag{nt}[c][c][1.2]{$c_{\mathtt{n}}[k]$}
\psfrag{rt}[c][c][1.2]{$r[k]$}
\psfrag{S}[c][c][1.2]{$\sum$}
\psfrag{P}[c][c][1.1]{$\mathrm{Poiss}$}
}}
\caption{Input-output signal model for an MC system with $L$ memory taps. Here, block ``$z^{-1}$" denotes one symbol interval delay and block ``$\mathrm{Poiss}$" generates a Poisson RV  with a  mean equal to the input \cite{Hamid_Lett,Adam_OptReciever}. The considered MC channel is characterized by $\bar{c}_l,\,\,l=1,\dots,L$,  i.e., the CIR. The goal of this paper is to estimate both the CIR and the mean of the noise, $\bar{c}_{\mathtt{n}}$,  based on RV $r[k]$.
 } 
\label{Fig:SysMod} 
\end{figure}

 Unlike the conventional linear input-output  model for channels with memory in wireless communication systems \cite{LSSE,Channel_Estimation_Arsalan}, the channel model in (\ref{Eq:ChannelInOut}) is not linear since $s[k-l+1]$ does not affect the observation $r[k]$ directly but via Poisson RV $c_l[k]$. However, the \textit{expectation} of the received signal is linearly dependent on the transmitted signal, i.e.,
\begin{IEEEeqnarray}{lll} \label{Eq:AveInOut}
 \bar{r}[k] = \mathbbmss{E}\left\{r[k]\right\} = \sum_{l=1}^{L} \bar{c}_l s[k-l+1] + \bar{c}_{\mathtt{n}}.
\end{IEEEeqnarray}
We note that for a given $s[k]$, in general, the actual number of molecules observed at the receiver, $r[k]$, will differ from the expected number of observed molecules, $\bar{r}[k]$, due to the intrinsic noisiness of diffusion. Here, we emphasize that the considered MC channel model is characterized by  $\bar{c}_l,\,\,l=1,\dots,L$, and $\bar{c}_{\mathtt{n}}$, see Fig.~\ref{Fig:SysMod}, which we refer to as the CIR of the MC channel throughout the the remainder of the paper.

\subsection{CIR Estimation Problem}

  Let $\mathbf{s}=[s[1],s[2],\dots,s[K]]^T$ be a training sequence of length $K$. Here, we assume continuous transmission. Therefore, in order to ensure that the received signal is only affected by training sequence $\mathbf{s}$ and not by the  transmissions in previous symbol intervals, we only employ $r[k],\,\,k\geq L$, for CIR estimation. Thereby, the $K-L+1$   samples  used for CIR estimation are given by 
\begin{IEEEeqnarray}{rll} \label{Eq:SysRecMol}
 r[L]   = & \mathrm{Poiss}\left(\bar{c}_1  s[L]\right) + \mathrm{Poiss}\left(\bar{c}_2  s[L-1]\right) + \cdots \nonumber \\
  & + \mathrm{Poiss}\left(\bar{c}_L  s[1]\right) + \mathrm{Poiss}\left(\bar{c}_{\mathtt{n}}\right)   \quad\,\,\,\, \IEEEyesnumber\IEEEyessubnumber \\
  r[L+1]   = & \mathrm{Poiss}\left(\bar{c}_1  s[L+1]\right) + \mathrm{Poiss}\left(\bar{c}_2  s[L]\right) + \cdots \nonumber \\ 
 & + \mathrm{Poiss}\left(\bar{c}_L  s[2]\right) + \mathrm{Poiss}\left(\bar{c}_{\mathtt{n}}\right)   \IEEEyessubnumber \\
  & \qquad\vdots   \qquad\qquad\qquad\qquad\qquad    \vdots  \nonumber \\
 r[K]   = & \mathrm{Poiss}\left(\bar{c}_1  s[K]\right) + \mathrm{Poiss}\left(\bar{c}_2  s[K-1]\right) + \cdots    \nonumber \\
  & + \mathrm{Poiss}\left(\bar{c}_L  s[K-L+1]\right) + \mathrm{Poiss}\left(\bar{c}_{\mathtt{n}}\right).   \qquad \IEEEyessubnumber 
\end{IEEEeqnarray}

For convenience of notation, we define $\mathbf{r}=[r[L],r[L+1],\dots,r[K]]^T$ and $\bar{\mathbf{c}}=[\bar{c}_1, \bar{c}_2,\dots,\bar{c}_L,\bar{c}_{\mathtt{n}}]^T$, and $f_{\mathbf{r}}(\mathbf{r}|\bar{\mathbf{c}},\mathbf{s})$ is the probability density function (PDF) of observation  $\mathbf{r}$  conditioned on a given channel $\bar{\mathbf{c}}$ and a given training sequence  $\mathbf{s}$. We assume that the CIR\footnote{With a slight abuse of notation,   we refer to vector $\bar{\mathbf{c}}$ as the CIR vector throughout the remainder of the paper although $\bar{\mathbf{c}}$ also contains the mean of the noise $\bar{c}_{\mathtt{n}}$.}, $\bar{\mathbf{c}}$, remains unchanged for a sufficiently large block of  symbol intervals during which CIR estimation and data transmission are performed. However, the CIR may change from one block to the next due to e.g. a change of the distance between transmitter and receiver. To model this, we assume that  CIR $\bar{\mathbf{c}}$ is a RV which takes its values in each block according to a PDF  $f_{\bar{\mathbf{c}}}(\bar{\mathbf{c}})$.  To summarize,  in each block, the stochastic model in (\ref{Eq:ChannelInOut}) is characterized by $\bar{\mathbf{c}}$ and  our goal in this paper is to estimate $\bar{\mathbf{c}}$ based on the vector of random observations $\mathbf{r}$.

\subsection{CR Lower Bound}

The classical CR bound is a lower bound on the variance of any unbiased estimator of a \textit{deterministic} parameter. However, this bound can be also generalized to stochastic parameters which leads to the Bayesian CR bound  \cite{BayesianBook,VanTreeEstDet}. In particular, under some regularity conditions\footnote{These conditions ensure that the Fisher information matrix exists. In particular, the continuous differentiablity of the PDFs $f_{\mathbf{r}}(\mathbf{r}|\bar{\mathbf{c}},\mathbf{s})$ and $f_{\mathbf{r},\bar{\mathbf{c}}}(\mathbf{r},\bar{\mathbf{c}}|\mathbf{s})$ is a sufficient condition~\cite{BayesianBook}.}, the covariance matrix of any unbiased estimate, $\hat{\bar{\mathbf{c}}}$, of parameter $\bar{\mathbf{c}}$, denoted by $\mathbf{C}(\hat{\bar{\mathbf{c}}})$, satisfies
\begin{IEEEeqnarray}{lll} \label{Eq:CRB_Defin}
\mathbf{C}\left(\hat{\bar{\mathbf{c}}} \right) - \mathbf{I}^{-1} \left(\bar{\mathbf{c}}\right) \succeq 0,
\end{IEEEeqnarray}
where  $\mathbf{I} \left( \bar{\mathbf{c}} \right)$ is the Fisher information matrix of parameter vector $\bar{\mathbf{c}} $ and is given by \cite{VanTreeEstDet}
\begin{IEEEeqnarray}{lll} \label{Eq:Fisher_Matrix}
\mathbf{I} \left( \bar{\mathbf{c}} \right)
= \begin{cases}
 \mathbbmss{E}_{\mathbf{r}|\bar{\mathbf{c}}}\left\{ - \nabla^2_{\bar{\mathbf{c}}\bar{\mathbf{c}}}  \mathrm{ln} \{f_{\mathbf{r}}(\mathbf{r}|\bar{\mathbf{c}},\mathbf{s})\}  \right\},   \\
 \qquad\qquad  \mathrm{if}\,\, \bar{\mathbf{c}} \text{ is a deterministic parameter} \\
 \mathbbmss{E}_{\mathbf{r},\bar{\mathbf{c}}}\left\{ - \nabla^2_{\bar{\mathbf{c}}\bar{\mathbf{c}}} \mathrm{ln} \{f_{\mathbf{r},\bar{\mathbf{c}}}(\mathbf{r},\bar{\mathbf{c}}|\mathbf{s})\}    \right\}, \\ 
 \qquad\qquad  \mathrm{if} \,\, \bar{\mathbf{c}} \text{ is a stochastic parameter}
\end{cases}
\end{IEEEeqnarray}
We assume that conditioned on $\bar{\mathbf{c}}$ and $\mathbf{s}$,  the observations in different symbol intervals are independent, i.e., $r[k]$ is independent of $r[k']$ for $k\neq k'$. This assumption is valid in practice if the time interval between two consecutive samples is sufficiently large, see \cite{Adam_OptReciever} for a detailed discussion. Moreover, from (\ref{Eq:ChannelInOut}), we observe that $r[k]$ is a sum of  Poisson RVs. Hence, $r[k]$ is also a Poisson RV with its mean equal to the sum of the means of the summands, i.e., $r[k]\sim\mathrm{Poiss}(\bar{r}[k])$ with $\bar{r}[k] = \bar{c}_{\mathtt{n}} + \sum_{l=1}^{L} \bar{c}_l s[k-l+1] = \bar{\mathbf{c}}^T\mathbf{s}_k  $ and $\mathbf{s}_k=[s[k],s[k-1],\dots,s[k-L+1],1]^T$.  Therefore, $f_{\mathbf{r}}(\mathbf{r}|\bar{\mathbf{c}},\mathbf{s})$ is given by
\begin{IEEEeqnarray}{rll} \label{Eq:ML_PDF}
 f_{\mathbf{r}}(\mathbf{r}|\bar{\mathbf{c}},\mathbf{s})\,\, 
& = \prod_{k=L}^{K} \frac{\left(\bar{\mathbf{c}}^T\mathbf{s}_k  \right)^{r[k]} \exp\left(-\bar{\mathbf{c}}^T\mathbf{s}_k  \right)}{r[k]!}. 
\end{IEEEeqnarray}

For a positive semidefinite matrix, the diagonal elements are non-negative, i.e., $\big[\mathbf{C}(\hat{\bar{\mathbf{c}}}) - \mathbf{I}^{-1} \left(\bar{\mathbf{c}} \right)\big]_{i,i} \geq 0$. Therefore, for an unbiased estimator, i.e., $\mathbbmss{E}\left\{\hat{\bar{\mathbf{c}}}\right\}=\bar{\mathbf{c}}$ holds, with the estimation error vector defined as $\mathbf{e} = \bar{\mathbf{c}} - \hat{\bar{\mathbf{c}}} $,  the classical CR bound for deterministic $\bar{\mathbf{c}}$ provides the following lower bound on the sum of the expected square errors
\begin{IEEEeqnarray}{lll} \label{Eq:CRB_CIR}
 \mathbbmss{E}_{\mathbf{r}|\bar{\mathbf{c}}} \left\{ \|\mathbf{e}\|^2\right\}  \\
  \geq \mathrm{tr}\left\{ \mathbf{I}^{-1} \left(\bar{\mathbf{c}}\right) \right\} 
 = \mathrm{tr}\left\{ \left[ \sum_{k=L}^{K} \frac{\mathbf{s}_k\mathbf{s}_k^T}{\bar{\mathbf{c}}^T \mathbf{s}_k} \right]^{-1} \right\} \triangleq \mathtt{CCRB}(\bar{\mathbf{c}}), \quad\nonumber
\end{IEEEeqnarray}
where $\mathtt{CCRB}(\bar{\mathbf{c}})$ denotes the classical CR bound for a given fixed/deterministic parameter $\bar{\mathbf{c}}$.

On the other hand, the Bayesian CR bound for stochastic $\bar{\mathbf{c}}$ is  given by
\begin{IEEEeqnarray}{lll} \label{Eq:CRB_CIR_Bays}
 \mathbbmss{E}_{\mathbf{r},\bar{\mathbf{c}}} \left\{ \|\mathbf{e}\|^2\right\}  \\ 
  \geq   \mathrm{tr}\left\{ \left[  \mathbbmss{E}_{\bar{\mathbf{c}}}\left\{ - \nabla^2_{\bar{\mathbf{c}}\bar{\mathbf{c}}}  \mathrm{ln} \{f_{\bar{\mathbf{c}}}(\bar{\mathbf{c}})\} + \sum_{k=L}^{K} \frac{\mathbf{s}_k\mathbf{s}_k^T}{\bar{\mathbf{c}}^T \mathbf{s}_k}  \right\} \right]^{-1} \right\}\triangleq \mathtt{BCRB}. \quad\nonumber
\end{IEEEeqnarray}
The expectation in (\ref{Eq:CRB_CIR_Bays}) is taken over RV  $\bar{\mathbf{c}}$ which takes its values according to PDF $f_{\bar{\mathbf{c}}}(\bar{\mathbf{c}})$. Note that $f_{\bar{\mathbf{c}}}(\bar{\mathbf{c}})$ depends on the MC environment  and a general analytical expression for $f_{\bar{\mathbf{c}}}(\bar{\mathbf{c}})$ is not yet   available in the literature. In practice,  $f_{\bar{\mathbf{c}}}(\bar{\mathbf{c}})$ can be obtained using  historical measurements of the CIR, see Remark~\ref{Remk:Fc} for further discussion. 

\begin{remk}
By comparing (\ref{Eq:CRB_CIR}) and (\ref{Eq:CRB_CIR_Bays}), we observe that $\mathbbmss{E}_{\bar{\mathbf{c}}} \left\{ \mathtt{CCRB}(\bar{\mathbf{c}}) \right\} \geq \mathtt{BCRB}$ holds as a result of Jensen's inequality \cite{Jensen}. In particular, for unbiased estimators which do not exploit any statistical knowledge of the MC channel, both lower bounds $\mathbbmss{E}_{\bar{\mathbf{c}}} \left\{ \mathtt{CCRB}(\bar{\mathbf{c}}) \right\}$ and $\mathtt{BCRB}$  are valid but $\mathbbmss{E}_{\bar{\mathbf{c}}} \left\{ \mathtt{CCRB}(\bar{\mathbf{c}}) \right\}$ is a tighter bound. On the other hand, for unbiased estimators which exploit some statistical knowledge of the MC channel,  $\mathbbmss{E}_{\bar{\mathbf{c}}} \left\{ \mathtt{CCRB}(\bar{\mathbf{c}}) \right\}$ is not a valid lower bound whereas $\mathtt{BCRB}$ is a valid lower bound. \hfill $\QED$
\end{remk}

\section{CIR Estimation without Statistical Channel Knowledge}

In this section, we consider estimators which do not exploit statistical knowledge of the MC channel. In particular, we derive the ML and the LSSE CIR estimators.

\subsection{ML CIR Estimation}

The ML CIR estimator finds the CIR estimate which maximizes the likelihood of  observation vector $\mathbf{r}$ \cite{BayesianBook}. In particular,  the ML estimator is given by
\begin{IEEEeqnarray}{lll} \label{Eq:ML_Estimation}
  \hat{\bar{\mathbf{c}}}^{\mathtt{ML}} = \underset{\bar{\mathbf{c}}\geq \mathbf{0}}{\mathrm{argmax}} \,\,f_{\mathbf{r}}(\mathbf{r}|\bar{\mathbf{c}},\mathbf{s}),
\end{IEEEeqnarray}
where $f_{\mathbf{r}}(\mathbf{r}|\bar{\mathbf{c}},\mathbf{s})$ is given in (\ref{Eq:ML_PDF}). Maximizing $f_{\mathbf{r}}(\mathbf{r}|\bar{\mathbf{c}},\mathbf{s})$ is equivalent to maximizing $\mathrm{ln}(f_{\mathbf{r}}(\mathbf{r}|\bar{\mathbf{c}},\mathbf{s}))$ since $\mathrm{ln}(\cdot)$ is a monotonically increasing function. Hence, the ML estimate  can be rewritten as
\begin{IEEEeqnarray}{cll} \label{Eq:ML_Log}
  \hat{\bar{\mathbf{c}}}^{\mathtt{ML}} = \underset{\bar{\mathbf{c}}\geq \mathbf{0}}{\mathrm{argmax}} \,\,g(\bar{\mathbf{c}}) \quad \mathrm{where}   \\ 
 g(\bar{\mathbf{c}}) \triangleq \sum_{k=L}^{K} \Big[-\bar{\mathbf{c}}^T\mathbf{s}_k   + r[k]\mathrm{ln}\left(\bar{\mathbf{c}}^T\mathbf{s}_k  \right)\Big].\nonumber
\end{IEEEeqnarray}
To present the solution of the above optimization problem rigorously, we first define some auxiliary variables. Let $\mathcal{A}=\{\mathcal{A}_1,\mathcal{A}_2,\dots,\mathcal{A}_N\}$ denote a set which contains all possible $N=2^{L+1}-1$ subsets of set $\mathcal{F}=\{1,2,\cdots,L,\mathtt{n}\}$ except for the empty set. Here, $\mathcal{A}_n,\,\,n=1,2,\dots,N$, denotes the $n$-th subset  of $\mathcal{A}$. Moreover, let $\bar{\mathbf{c}}^{\mathcal{A}_n}$ and $\mathbf{s}_k^{\mathcal{A}_n}$ denote reduced-dimension versions of  $\bar{\mathbf{c}}$ and $\mathbf{s}_k$, respectively, which contain only those elements of $\bar{\mathbf{c}}$ and $\mathbf{s}_k$ whose indices are elements of set $\mathcal{A}_n$, respectively.

\begin{lem}\label{Lem:ML}
The ML estimator of the CIR for the considered MC channel is given by Algorithm~1 where the following non-linear system of equations is solved\footnote{A system of nonlinear equations can be solved using mathematical software packages, e.g. Mathematica.} for different $\mathcal{A}_n$
\begin{IEEEeqnarray}{lll} \label{Eq:ML_Sol}
\sum_{k=L}^{K} \left[\frac{r[k] }{(\bar{\mathbf{c}}^{\mathcal{A}_n})^T \mathbf{s}_k^{\mathcal{A}_n}} -1\right]  \mathbf{s}_k^{\mathcal{A}_n} = \mathbf{0}.
\end{IEEEeqnarray}
\end{lem}

\begin{IEEEproof}
Please refer to Appendix~\ref{App:ML}.
\end{IEEEproof}

\begin{algorithm}[t] 
\caption{  
{\color[rgb]{0,0,1}ML}/{\color[rgb]{1,0,0}LSSE} CIR Estimate {\color[rgb]{0,0,1}$\hat{\bar{\mathbf{c}}}^{\mathtt{ML}}$}/{\color[rgb]{1,0,0}$\hat{\bar{\mathbf{c}}}^{\mathtt{LSSE}}$}
  }
\begin{algorithmic} 
\STATE \textbf{initialize} $\mathcal{A}_n=\mathcal{F}$ and solve {\color[rgb]{0,0,1}(\ref{Eq:ML_Sol})}/{\color[rgb]{1,0,0}(\ref{Eq:LSSE_Sol})}  to find $\bar{\mathbf{c}}^{\mathcal{F}}$ 
\IF{$\bar{\mathbf{c}}^{\mathcal{F}}\geq \mathbf{0}$}
\STATE  Set {\color[rgb]{0,0,1}$\hat{\bar{\mathbf{c}}}^{\mathtt{ML}} = \bar{\mathbf{c}}^{\mathcal{F}}$}/{\color[rgb]{1,0,0}$\hat{\bar{\mathbf{c}}}^{\mathtt{LSSE}} = \bar{\mathbf{c}}^{\mathcal{F}}$}
\ELSE
 \FOR{$\forall \mathcal{A}_n\neq \mathcal{F}$}
        \STATE Solve {\color[rgb]{0,0,1}(\ref{Eq:ML_Sol})}/{\color[rgb]{1,0,0}(\ref{Eq:LSSE_Sol})} to find $\bar{\mathbf{c}}^{\mathcal{A}_n}$ 
        \IF{$\bar{\mathbf{c}}^{\mathcal{A}_n}\geq\mathbf{0}$ holds}
        \STATE Set the values of the elements of $\hat{\bar{\mathbf{c}}}^{\mathtt{CAN}}$, whose indices are in $\mathcal{A}_n$, equal to the values of the corresponding elements in $\bar{\mathbf{c}}^{\mathcal{A}_n}$ and the remaining $L+1-\left|\mathcal{A}_n\right|$ elements equal to zero;
        \STATE Save $\hat{\bar{\mathbf{c}}}^{\mathtt{CAN}}$ in the candidate set $\mathcal{C}$
        \ELSE
        \STATE Discard $\bar{\mathbf{c}}^{\mathcal{A}_n}$
        \ENDIF
 \ENDFOR
 \STATE Choose {\color[rgb]{0,0,1}$\hat{\bar{\mathbf{c}}}^{\mathtt{ML}}$}/{\color[rgb]{1,0,0}$\hat{\bar{\mathbf{c}}}^{\mathtt{LSSE}}$} equal to that $\hat{\bar{\mathbf{c}}}^{\mathtt{CAN}}$ in the candidate set $\mathcal{C}$ which  {\color[rgb]{0,0,1}maximizes $g(\bar{\mathbf{c}})$}/{\color[rgb]{1,0,0}minimizes $\|\boldsymbol{\epsilon}\|^2$}
\ENDIF
\end{algorithmic}
\end{algorithm}

\begin{remk}\label{Remk:ML_Unbiased}
Recall  that in the system
model, we  assumed  a priori $L$ non-zero taps and a noise with non-zero mean, i.e., $\bar{\mathbf{c}} > \mathbf{0}$. Moreover, the consistency property of ML estimation \cite[Chapter 4]{BayesianBook} implies that under some regularity
conditions, notably that the likelihood is a continuous function of $\bar{\mathbf{c}}$ and that $\bar{\mathbf{c}}$ is not on the boundary of the parameter set $\bar{\mathbf{c}} \geq \mathbf{0}$, we obtain $\hat{\bar{\mathbf{c}}}^{\mathtt{ML}}\to\bar{\mathbf{c}}$ as $K\to\infty$. As a result, the ML estimator is asymptotically unbiased.  Therefore, for large values of $K$, the ML estimator  becomes  sufficiently  accurate  such  that  none  of  the elements  of  $\hat{\bar{\mathbf{c}}}^{\mathtt{ML}}$ is  zero.  In  this  case,  Algorithm  1  reduces  to
directly solving (\ref{Eq:ML_Sol}) for $\mathcal{A}_n=\mathcal{F}$,  where the left-hand side of (\ref{Eq:ML_Sol}) is the derivative of $g(\bar{\mathbf{c}})$ given in (\ref{Eq:ML_Log}) with respect to $\bar{\mathbf{c}}$.  \hfill$\QED$
\end{remk}

Motivated by the results in Lemma~\ref{Lem:ML} and the discussion in Remark~\ref{Remk:ML_Unbiased}, we propose the following simpler but suboptimal estimate based on the ML estimate in (\ref{Eq:ML_Sol}). 

\textit{Suboptimal ML-Based CIR Estimation:} For a given observation vector $\mathbf{r}$, the suboptimal estimate $\hat{\bar{\mathbf{c}}}^{\mathtt{ML}}_{\mathtt{sub}}$ is given by $\hat{\bar{\mathbf{c}}}^{\mathtt{ML}}_{\mathtt{sub}}= \left[\bar{\mathbf{c}}\right]^+$ where $\bar{\mathbf{c}}$ is the solution of the following system of equations
\begin{IEEEeqnarray}{lll} \label{Eq:ML_Sub}
 \sum_{k=L}^{K} \left[\frac{r[k] }{\bar{\mathbf{c}}^T \mathbf{s}_k } -1\right]  \mathbf{s}_k  = \mathbf{0}.
\end{IEEEeqnarray}
Note that the above estimate is simpler than the optimal ML estimate in Lemma~\ref{Lem:ML} since in the case that the $\bar{\mathbf{c}}$ found from (\ref{Eq:ML_Sub}) does not satisfy $\bar{\mathbf{c}} \geq \mathbf{0}$, we do not search for the optimal boundary points and instead just set the negative elements of $\bar{\mathbf{c}}$ to zero. Moreover, considering Remark~\ref{Remk:ML_Unbiased}, this simpler estimate becomes asymptotically identical to the ML estimate, i.e., as $K\to\infty$, we obtain $\hat{\bar{\mathbf{c}}}^{\mathtt{ML}}_{\mathtt{sub}}\to\hat{\bar{\mathbf{c}}}^{\mathtt{ML}}$. As we will see  in Section~VI, cf. Fig.~\ref{Fig:Var_K}, $\hat{\bar{\mathbf{c}}}^{\mathtt{ML}}_{\mathtt{sub}}$ approaches the performance of $\hat{\bar{\mathbf{c}}}^{\mathtt{ML}}$ even  for small $K$.

\begin{remk}
We note that obtaining the ML estimate of the CIR might be computationally challenging for practical MC systems as nanonodes have limited computational power. Nevertheless,  ML estimation can serve as a benchmark for low-complexity estimators such as the LSSE estimator. Moreover, for applications where the nanoreceiver only  collects observations, i.e., vector $\mathbf{r}$, and forwards them to an external processing unit outside the MC environment, the computational complexity of ML estimation may be affordable. This case may apply e.g. in health monitoring when a computer outside the body may be available for offline processing. \hfill$\QED$
\end{remk}

\subsection{LSSE CIR Estimation}

The LSSE CIR estimator chooses that $\bar{\mathbf{c}}$ which minimizes the sum of the square errors for observation vector $\mathbf{r}$. Thereby, the error vector is defined as $\boldsymbol{\epsilon} = \mathbf{r} - \mathbbmss{E}\left\{\mathbf{r}\right\} = \mathbf{r} - \mathbf{S} \bar{\mathbf{c}}$ where $\mathbf{S} = [\mathbf{s}_L,\mathbf{s}_{L+1},\dots,\mathbf{s}_K]^T$. In particular, the LSSE CIR estimate can be formally written as
\begin{IEEEeqnarray}{lll} \label{Eq:LSSE_Estimation}
\hat{\bar{\mathbf{c}}}^{\mathtt{LSSE}} = \underset{\bar{\mathbf{c}}\geq \mathbf{0}}{\mathrm{argmin}} \,\, \| \boldsymbol{\epsilon} \|^2.
\end{IEEEeqnarray}
Here, the square norm of the  error vector is obtained as
\begin{IEEEeqnarray}{lll} \label{Eq:LSSE_ErrorNorm}
 \|\boldsymbol{\epsilon}\|^2 & = \mathrm{tr}\left\{ \boldsymbol{\epsilon}\boldsymbol{\epsilon}^T \right\}=  \mathrm{tr}\left\{ (\mathbf{r} - \mathbf{S} \bar{\mathbf{c}}) (\mathbf{r} - \mathbf{S} \bar{\mathbf{c}})^T\right\} \nonumber \\
  &= \mathrm{tr}\left\{ \mathbf{S}^T \mathbf{S} \bar{\mathbf{c}} \bar{\mathbf{c}}^T  \right\} - 2 \mathrm{tr}\left\{ \mathbf{r}^T\mathbf{S}\bar{\mathbf{c}} \right\} + \mathrm{tr}\left\{ \mathbf{r} \mathbf{r}^T \right\},
\end{IEEEeqnarray}
where we used the following properties of the trace:  $\mathrm{tr}\left\{\mathbf{A}\right\}=\mathrm{tr}\left\{\mathbf{A}^T\right\}$ and $\mathrm{tr}\left\{\mathbf{A}\mathbf{B}\right\}=\mathrm{tr}\left\{\mathbf{BA}\right\}$ \cite{TraceDerivative}. The LSSE estimate is given in the following lemma where we use the auxiliary matrix  $\mathbf{S}^{\mathcal{A}_n} = [\mathbf{s}_L^{\mathcal{A}_n} ,\mathbf{s}_{L+1}^{\mathcal{A}_n} ,\dots,\mathbf{s}_K^{\mathcal{A}_n} ]^T$.

\begin{lem}\label{Lem:LSSE}
The LSSE estimator of the CIR for the considered MC channel is given by Algorithm~1 where for a given set $\mathcal{A}_n$,  $\bar{\mathbf{c}}^{\mathcal{A}_n}$ is obtained as  
\begin{IEEEeqnarray}{lll} \label{Eq:LSSE_Sol}
  \bar{\mathbf{c}}^{\mathcal{A}_n} = \left((\mathbf{S}^{\mathcal{A}_n} )^T \mathbf{S}^{\mathcal{A}_n} \right)^{-1}  (\mathbf{S}^{\mathcal{A}_n})^T \mathbf{r}.
\end{IEEEeqnarray}
\end{lem}
\begin{IEEEproof}
  Please refer to Appendix~\ref{App:LSSE}.
\end{IEEEproof}

 \begin{remk}
The LSSE estimator employs in fact a linear matrix multiplication to compute $\bar{\mathbf{c}}^{\mathcal{A}_n}$, i.e., $\bar{\mathbf{c}}^{\mathcal{A}_n} = \mathbf{F}^{\mathcal{A}_n}\mathbf{r}$ where $\mathbf{F}^{\mathcal{A}_n}= \left((\mathbf{S}^{\mathcal{A}_n} )^T \mathbf{S}^{\mathcal{A}_n} \right)^{-1}  (\mathbf{S}^{\mathcal{A}_n})^T$. Moreover, since the training sequence $\mathbf{s}$ is fixed,  matrix  $\mathbf{F}^{\mathcal{A}_n}$ can be calculated offline and then be used  for online CIR estimation. Therefore, the calculation of  $\hat{\bar{\mathbf{c}}}^{\mathcal{A}_n}$ for the LSSE estimator in (\ref{Eq:LSSE_Sol}) is considerably less computationally complex than the computation of $\hat{\bar{\mathbf{c}}}^{\mathcal{A}_n}$ for the ML estimator  in (\ref{Eq:ML_Sol}) which requires solving a system of nonlinear  equations. \hfill$\QED$
\end{remk}

 \begin{corol}\label{Corol:LSSE_Unbiased}
The LSSE estimator in Lemma~\ref{Lem:LSSE} is biased in general but asymptotically, as $K\to\infty$, becomes unbiased. More precisely, the LSSE estimator in Lemma~\ref{Lem:LSSE} is a consistent estimator, i.e., $\hat{\bar{\mathbf{c}}}^{\mathtt{LSSE}}\to\bar{\mathbf{c}}$ as $K\to\infty$.
\end{corol} 
\begin{IEEEproof}
 Please refer to Appendix~\ref{App:LSSE_Consis}.
\end{IEEEproof}

The above corollary reveals that similar to the ML estimate, the LSSE estimate is biased for short length sequences but becomes asymptotically unbiased as the sequence length grows to infinity. Moreover, motivated by the results in Corollary~\ref{Corol:LSSE_Unbiased} and similar to the sub-optimal estimator proposed based on the ML estimate, we propose the following suboptimal estimator based on the LSSE estimate in (\ref{Eq:LSSE_Sol}). 

\textit{Suboptimal LSSE-Based CIR Estimation:} For a given observation vector $\mathbf{r}$, the suboptimal estimate $\hat{\bar{\mathbf{c}}}^{\mathtt{LSSE}}_{\mathtt{sub}}$ is given by 
\begin{IEEEeqnarray}{lll} \label{Eq:LSSE_Sub}
\hat{\bar{\mathbf{c}}}^{\mathtt{LSSE}}_{\mathtt{sub}}=\left[\left(\mathbf{S}^T \mathbf{S}\right)^{-1}  \mathbf{S}^T \mathbf{r}\right]^+.
\end{IEEEeqnarray}
Note that the above suboptimal estimate  asymptotically, as $K\to\infty$, converges to the LSSE estimate, i.e., $\hat{\bar{\mathbf{c}}}^{\mathtt{LSSE}}_{\mathtt{sub}}\to\hat{\bar{\mathbf{c}}}^{\mathtt{LSSE}}$. Moreover, as we will see in Section~VI, cf. Fig.~\ref{Fig:Var_K}, $\hat{\bar{\mathbf{c}}}^{\mathtt{LSSE}}_{\mathtt{sub}}$ approaches the performance of $\hat{\bar{\mathbf{c}}}^{\mathtt{LSSE}}$ even for small $K$.

\begin{remk}\label{Remk:ML_LSSE_AsymUnbiased}
We note that for finite $K$, the ML and LSSE estimators in Algorithm~1 are biased in general. The proposed estimators are biased because of the non-negativity of the unknown parameters. Exploiting this information in the proposed estimators causes the biasedness for small $K$ but improves the estimation performance by reducing the variance of the estimation error. Hence, the error variances of the ML and LSSE estimates  may fall below the classical CR bound. However,  as $K\to\infty$, the ML and LSSE estimators become asymptotically unbiased, cf. Remark~\ref{Remk:ML_Unbiased} and Corollary~\ref{Corol:LSSE_Unbiased}, and the classical CR  bound becomes a valid lower bound. The asymptotic unbiasedness   of the proposed estimators is also numerically verified in Section~VI, cf. Fig.~\ref{Fig:Mean_K}. \hfill$\QED$
\end{remk}

\begin{remk} The appropriate length of the training sequence depends on the required CIR accuracy and the coherence time of the MC channel. In particular, shorter training sequences can be used if higher estimation errors can be tolerated. Furthermore, since the coherence time of the MC channel determines how fast the MC channel changes, the length of the training sequence should be chosen small compared to the coherence time such that the  CIR acquisition overhead is low. \hfill$\QED$
\end{remk}

\section{CIR Estimation with Statistical Channel Knowledge}

In this section, we first present the MAP  estimator assuming that full statistical knowledge of the MC channel is available. Subsequently, we derive the LMMSE  estimator for the case when only the first and second order statistics of the MC channel are available.

\subsection{MAP CIR Estimation}

The MAP CIR estimator is the optimal estimator assuming that full statistical knowledge of the MC channel is available, i.e.,   $f_{\bar{\mathbf{c}}}(\bar{\mathbf{c}})$. In particular,  the MAP CIR estimator chooses that CIR $\bar{\mathbf{c}}$ which maximizes the posterior expected value of observation vector $\mathbf{r}$, i.e., $f_{\bar{\mathbf{c}}}(\bar{\mathbf{c}}|\mathbf{r},\mathbf{s})$. Thereby, the MAP estimate is given by 
\begin{IEEEeqnarray}{lll} \label{Eq:MAP_Estimation}
  \hat{\bar{\mathbf{c}}}^{\mathtt{MAP}} = \underset{\bar{\mathbf{c}}\geq \mathbf{0}}{\mathrm{argmax}} \,\,f_{\bar{\mathbf{c}}}(\bar{\mathbf{c}}|\mathbf{r},\mathbf{s})
  \overset{(a)}{=}  \frac{f_{\mathbf{r}}(\mathbf{r}|\bar{\mathbf{c}},\mathbf{s})f_{\bar{\mathbf{c}}}(\bar{\mathbf{c}})}{f_{\mathbf{r}}(\mathbf{r})},
\end{IEEEeqnarray}
where $(a)$ follows the Bayes' Theorem \cite{BayesianBook}. Note that the term $f_{\mathbf{r}}(\mathbf{r})$ does not depend on the estimate $\bar{\mathbf{c}}$.  Moreover, considering that $\mathrm{ln}(\cdot)$ is a monotonically increasing function, the MAP estimate  can be rewritten in the following convenient form
\begin{IEEEeqnarray}{cll} \label{Eq:MAP_Log}
  \hat{\bar{\mathbf{c}}}^{\mathtt{MAP}} = \underset{\bar{\mathbf{c}}\geq \mathbf{0}}{\mathrm{argmax}} \,\,\left[ g(\bar{\mathbf{c}}) + \mathrm{ln}\left( f_{\bar{\mathbf{c}}}(\bar{\mathbf{c}}) \right)\right],
\end{IEEEeqnarray}
where $g(\bar{\mathbf{c}})$ is given in (\ref{Eq:ML_Log}).

\begin{remk}\label{Remk:Fc}
The PDF $f_{\bar{\mathbf{c}}}(\bar{\mathbf{c}})$ depends on the considered MC environment  and  a general analytical expression for $f_{\bar{\mathbf{c}}}(\bar{\mathbf{c}})$ is not yet  available in the literature.  In practice,  $f_{\bar{\mathbf{c}}}(\bar{\mathbf{c}})$ for a particular MC channel can be obtained using  historical measurements of $\bar{\mathbf{c}}$. Another convenient approximation for mathematical derivations is to assume a particular shape for $f_{\bar{\mathbf{c}}}(\bar{\mathbf{c}})$ and adjust the parameters of the PDF to match the experimental data. For the simulation results presented in Section~VI, we assume that  $f_{\bar{\mathbf{c}}}(\bar{\mathbf{c}})$ is a multivariant Gaussian PDF\footnote{We note that $\bar{\mathbf{c}}\geq \mathbf{0}$ has to hold, whereas a Gaussian PDF may lead to negative-valued realizations for the elements of $\bar{\mathbf{c}}$. To resolve this issue, in Section~VI, we use $\bar{\mathbf{c}}=[\bar{\mathbf{c}}^{\mathtt{Gaus}}]^+$, where $\bar{\mathbf{c}}^{\mathtt{Gaus}}$ is a Gaussian distributed vector. Thereby, for mathematical tractability of computing $ \hat{\bar{\mathbf{c}}}^{\mathtt{MAP}}$, we adopt the Gaussian PDF $f_{\bar{\mathbf{c}}^{\mathtt{Gaus}}}(\bar{\mathbf{c}}^{\mathtt{Gaus}})$ as an approximation for $f_{\bar{\mathbf{c}}}(\bar{\mathbf{c}})$.} with a certain mean vector $\boldsymbol{\mu}_{\bar{\mathbf{c}}}$ and covariance matrix $\boldsymbol{\Phi}_{\bar{\mathbf{c}}\bar{\mathbf{c}}}$ which are chosen based on our simulation parameters. However, we emphasize that the choice of the CIR distribution depends on the specific MC channel. Here, as an example, we adopt the Gaussian distribution for the PDF of the CIR.  \hfill$\QED$
\end{remk}

Similar to the ML and LSSE estimates provided in the previous section, the optimal MAP estimate is either a stationary point or a boundary point, see Algorithm~1. Here, we do not provide a detailed solution for the MAP estimate due to space constraints, instead, we highlight the main steps for the case when the solution is a stationary point. For the case when the solution is a boundary point, we can employ the same methodology as in Algorithm~1 to find $\hat{\bar{\mathbf{c}}}^{\mathtt{MAP}}$. In particular, the stationary points of the cost function in (\ref{Eq:MAP_Log}) are obtained by equating its derivative to zero, i.e., $\frac{\partial g(\bar{\mathbf{c}}) }{\partial \bar{\mathbf{c}}} + \frac{\partial \mathrm{ln}\left( f_{\bar{\mathbf{c}}}(\bar{\mathbf{c}}) \right) }{\partial \bar{\mathbf{c}}} = \mathbf{0}$. Note that the first term $\frac{\partial g(\bar{\mathbf{c}}) }{\partial \bar{\mathbf{c}}}$ is given on the left-hand side of (\ref{Eq:ML_Sol}) for $\mathcal{A}_n=\mathcal{F}$ and the second term is given by $\frac{\partial \mathrm{ln}\left( f_{\bar{\mathbf{c}}}(\bar{\mathbf{c}}) \right) }{\partial \bar{\mathbf{c}}} = -\frac{1}{2} \left( \boldsymbol{\Phi}_{\bar{\mathbf{c}}\bar{\mathbf{c}}}^{-1}+\boldsymbol{\Phi}_{\bar{\mathbf{c}}\bar{\mathbf{c}}}^{-T}\right)(\bar{\mathbf{c}}-\boldsymbol{\mu}_{\bar{\mathbf{c}}})$ for the considered Gaussian PDF $f_{\bar{\mathbf{c}}}(\bar{\mathbf{c}})$. We note that although the complexity of the MAP estimator may not be affordable for practical MC systems, we can still use it as a benchmark scheme to evaluate the performance of less complex estimators, e.g., the LMMSE estimator.

\subsection{LMMSE CIR Estimation}

The MAP estimator in (\ref{Eq:MAP_Log}) requires full knowledge of the channel statistics, i.e., $f_{\bar{\mathbf{c}}}(\bar{\mathbf{c}})$, which is difficult to obtain for practical MC channels, cf Remark~\ref{Remk:Fc}. Therefore, in the following, we consider an LMMSE-based estimator which requires only knowledge of the first and second order statistics of the MC channel. In particular, in estimation theory, the average square norm of the estimation error  $\mathbf{e} = \bar{\mathbf{c}} - \hat{\bar{\mathbf{c}}}$ is often adopted as performance metric, i.e., $\mathbbmss{E}\left\{\|\mathbf{e}\|^2\right\}$. Therefore, one can design an LMMSE estimator  which directly minimizes $\mathbbmss{E}\left\{\|\mathbf{e}\|^2\right\}$. We note that the estimation error has to be averaged over both $\mathbf{r}$ and $\bar{\mathbf{c}}$, i.e.,  $\mathbbmss{E}_{\mathbf{r},\bar{\mathbf{c}}}\left\{\|\mathbf{e}\|^2\right\}$, since  both  are randomly changing, and hence, have to be modeled as RVs.   In this paper, we consider an LMMSE-based estimator given by $\hat{\bar{\mathbf{c}}}^{\mathtt{MMSE}} = \left[\mathbf{F}^{\mathtt{MMSE}}\mathbf{r}\right]^+$ where $\mathbf{F}^{\mathtt{MMSE}}$ is the MMSE matrix.   Moreover, for mathematical tractability, we optimize $\mathbf{F}^{\mathtt{MMSE}}$ such that an upper bound on the estimation error is minimized, i.e., 
\begin{IEEEeqnarray}{ccc} \label{Eq:MMSE_Estimation}
\mathbf{F}^{\mathtt{MMSE}}= \underset{\mathbf{F}}{\mathrm{argmin}} \,\, \mathbbmss{E}_{\mathbf{r},\bar{\mathbf{c}}}\left\{\|\mathbf{e}^{\mathtt{MMSE}}_{\mathtt{up}}\|^2\right\},
\end{IEEEeqnarray}
where $\mathbf{e}^{\mathtt{MMSE}}_{\mathtt{up}} = \bar{\mathbf{c}} - \hat{\bar{\mathbf{c}}}^{\mathtt{MMSE}}_{\mathtt{up}}$ and $\hat{\bar{\mathbf{c}}}^{\mathtt{MMSE}}_{\mathtt{up}} = \mathbf{F}^{\mathtt{MMSE}}\mathbf{r}$. Note that $\hat{\bar{\mathbf{c}}}^{\mathtt{MMSE}}_{\mathtt{up}}$ yields an upper bound on the estimation error of $\hat{\bar{\mathbf{c}}}^{\mathtt{MMSE}}$ because we neglect the side information that the CIR coefficients have to be non-negative by dropping $[\cdot]^+$. Hereby, $\mathbbmss{E}_{\mathbf{r},\bar{\mathbf{c}}}\left\{\|\mathbf{e}^{\mathtt{MMSE}}_{\mathtt{up}}\|^2\right\}$ is obtained as
\begin{IEEEeqnarray}{lll} \label{Eq:MMSE_ErrorNorm}
\mathbbmss{E}_{\mathbf{r},\bar{\mathbf{c}}}\left\{\|\mathbf{e}^{\mathtt{MMSE}}_{\mathtt{up}}\|^2\right\} 
= \mathbbmss{E}_{\mathbf{r},\bar{\mathbf{c}}}\left\{\mathrm{tr}\left\{\mathbf{e}^{\mathtt{MMSE}}_{\mathtt{up}}\left(\mathbf{e}^{\mathtt{MMSE}}_{\mathtt{up}}\right)^T \right\}\right\}   \nonumber \\ 
\qquad = \mathbbmss{E}_{\mathbf{r},\bar{\mathbf{c}}}\left\{\mathrm{tr}\left\{(\mathbf{F}\mathbf{r}-\bar{\mathbf{c}} )(\mathbf{F}\mathbf{r}-\bar{\mathbf{c}} )^T\right\}\right\} \nonumber \\
\qquad = \mathrm{tr}\left\{\mathbf{F} \boldsymbol{\Phi}_\mathbf{rr} \mathbf{F}^T - \mathbf{F} \boldsymbol{\Phi}_{\mathbf{r}\bar{\mathbf{c}}} 
    - \boldsymbol{\Phi}_{\bar{\mathbf{c}}\mathbf{r}} \mathbf{F}^T + \boldsymbol{\Phi}_{\bar{\mathbf{c}}\bar{\mathbf{c}}} \right\}, \quad
\end{IEEEeqnarray}
where
\begin{IEEEeqnarray}{lll} \label{Eq:MMSE_Qvar}
 \boldsymbol{\Phi}_\mathbf{rr} &=  \mathbbmss{E}_{\bar{\mathbf{c}}}\mathbbmss{E}_{\mathbf{r}|\bar{\mathbf{c}}}\left\{ \mathbf{r}\mathbf{r}^T\right\} 
 = \mathbbmss{E}_{\bar{\mathbf{c}}} \left\{ \mathbf{S}\bar{\mathbf{c}}\bar{\mathbf{c}}^T \mathbf{S}^T + \mathrm{diag}\left\{\mathbf{S}\bar{\mathbf{c}}\right\} \right\} \nonumber\\ 
&= \mathbf{S}\boldsymbol{\Phi}_{\bar{\mathbf{c}}\bar{\mathbf{c}}} \mathbf{S}^T + \mathrm{diag}\left\{\mathbf{S}\boldsymbol{\mu}_{\bar{\mathbf{c}}}\right\} \quad \IEEEyesnumber \IEEEyessubnumber \\
 \boldsymbol{\Phi}_{\mathbf{r}\bar{\mathbf{c}}} &= \boldsymbol{\Phi}_{\bar{\mathbf{c}}\mathbf{r}}^T = \mathbbmss{E}_{\bar{\mathbf{c}}}\mathbbmss{E}_{\mathbf{r}|\bar{\mathbf{c}}}\left\{ \mathbf{r}\bar{\mathbf{c}}^T\right\} = \mathbbmss{E}_{\bar{\mathbf{c}}} \left\{ \mathbf{S} \bar{\mathbf{c}}\bar{\mathbf{c}}^T\right\} = \mathbf{S}\boldsymbol{\Phi}_{\bar{\mathbf{c}}\bar{\mathbf{c}}}, \quad \IEEEyessubnumber  
\end{IEEEeqnarray}
and $\boldsymbol{\mu}_{\bar{\mathbf{c}}} = \mathbbmss{E}_{\bar{\mathbf{c}}}\{\bar{\mathbf{c}}\}$ and $\boldsymbol{\Phi}_{\bar{\mathbf{c}}\bar{\mathbf{c}}} =  \mathbbmss{E}_{\bar{\mathbf{c}}} \left\{  \bar{\mathbf{c}}\bar{\mathbf{c}}^T\right\}$. In addition, in (\ref{Eq:MMSE_Qvar}a) and (\ref{Eq:MMSE_Qvar}b), we use  $\mathbbmss{E}_{\mathbf{X}}\left\{\mathrm{tr}\left\{\mathbf{AXB}\right\}\right\} = \mathrm{tr}\left\{\mathbf{A}\mathbbmss{E}_{\mathbf{X}}\left\{\mathbf{X}\right\}\mathbf{B}\right\}$, which holds for  general matrices $\mathbf{A}$, $\mathbf{B}$, and $\mathbf{X}$. Furthermore,  in (\ref{Eq:MMSE_Qvar}a), we use  $\mathbbmss{E}_{\mathbf{x}}\left\{\mathbf{x}\mathbf{x}^T\right\}  = \boldsymbol{\lambda}\boldsymbol{\lambda}^T + \mathrm{diag}\{\boldsymbol{\lambda}\}$, which is valid for  multivariate Poisson random vectors $\mathbf{x}$ with covariance matrix $\mathbf{C}(\mathbf{x}) = \mathrm{diag}\{\boldsymbol{\lambda}\}$. The optimal MMSE matrix is given in the following lemma.

\begin{lem}\label{Lem:MMSE}
For the LMMSE-based estimator  $\hat{\bar{\mathbf{c}}}^{\mathtt{MMSE}} = \left[\mathbf{F}^{\mathtt{MMSE}}\mathbf{r}\right]^+$, the optimal matrix $\mathbf{F}^{\mathtt{MMSE}}$, i.e., the solution of the optimization problem in (\ref{Eq:MMSE_Estimation}), is given by 
\begin{IEEEeqnarray}{lll} \label{Eq:MMSE_Sol}
 \mathbf{F}^{\mathtt{MMSE}} = \boldsymbol{\Phi}_{\bar{\mathbf{c}}\bar{\mathbf{c}}}^T \mathbf{S}^T  \left(\mathbf{S}\boldsymbol{\Phi}_{\bar{\mathbf{c}}\bar{\mathbf{c}}} \mathbf{S}^T +  \mathrm{diag}\left\{\mathbf{S}\boldsymbol{\mu}_{\bar{\mathbf{c}}}\right\} \right)^{-1}. 
\end{IEEEeqnarray}
\end{lem}
\begin{IEEEproof}
Please refer to Appendix~\ref{App:MMSE}.
\end{IEEEproof}

\begin{remk}
Note that  the optimal MMSE matrix depends only on the training sequence and the first and second order channel statistics. Thus, the MMSE matrix can be computed offline once and then be used for online CIR acquisition as long as the channel statistics remain constant. Since the statistics of the CIR, $\bar{\mathbf{c}}$, change much less frequently than the CIR itself, an infrequent update of the MMSE matrix is sufficient. As we will see in Section~VI,  the LMMSE estimator provides a considerable performance gain compared to the ML and LSSE estimators at the cost of acquiring the required knowledge of the MC channel statistics. \hfill$\QED$
\end{remk}

\begin{remk}\label{Remk:MAP_MMSE_Unbiased}
Similar to the discussion  regarding the ML and LSSE estimators in Remark~\ref{Remk:ML_LSSE_AsymUnbiased}, we note that the proposed MAP and LMMSE estimators are in general biased for short sequence lengths and become asymptotically unbiased as $K\to\infty$. Therefore, the Bayesian CR bound in (\ref{Eq:CRB_CIR_Bays}) may not be a valid lower bound for the MAP and LMMSE estimators for small values of $K$, however, for $K\to\infty$, it becomes asymptotically a valid lower bound. Due to space constraints, we skip the detailed proofs for the biasedness and asymptotic unbiasedness of the proposed MAP and LMMSE estimators. Instead, we numerically illustrate these properties in Section~VI, cf. Fig.~\ref{Fig:Mean_K}.   \hfill$\QED$
\end{remk}

\section{Training Sequence Design}

In this section, we first present two optimal training sequence designs  for CIR estimation in MC systems based on the LSSE and LMMSE estimators.  Subsequently, we propose a suboptimal and simple training sequence which is suitable for practical applications where the involved nanomachines have limited computational processing capability. For future reference, let $\mathcal{S}$ and $\mathcal{S}^K$ denote the sets of all CSK symbols and all possible training sequences, respectively. 

\subsection{LSSE-Based Training Sequence Design}

We first consider a  training sequence design which  minimizes an \textit{upper bound} on the \textit{average} estimation error of the LSSE estimator.  First, we note that for training sequence design, the estimation error has to be averaged over both $\mathbf{r}$ and $\bar{\mathbf{c}}$ since both  are a priori unknown, and hence, have to be modeled as RVs.  Again, we  recall   that in the system model, we  assumed  a priori $L$ non-zero taps and a noise with non-zero mean. Therefore, neglecting the information that  $\bar{\mathbf{c}} \geq \mathbf{0}$ has to hold in (\ref{Eq:LSSE_Estimation}) yields an upper bound on the estimation error for the LSSE estimator. This upper bound is adopted here for the problem of sequence design since the solution of (\ref{Eq:LSSE_Estimation}) after dropping constraint $\bar{\mathbf{c}} \geq \mathbf{0}$ leads to the simple and elegant closed-form solution $\hat{\bar{\mathbf{c}}}^{\mathtt{LSSE}}_{\mathtt{up}}=\left(\mathbf{S}^T \mathbf{S}\right)^{-1}  \mathbf{S} \mathbf{r}$,  which we can employ to find optimal training sequences.  Moreover, this upper bound is tight as $K\to\infty$ since $\hat{\bar{\mathbf{c}}}^{\mathtt{LSSE}}>\mathbf{0}$ holds and we obtain $\hat{\bar{\mathbf{c}}}^{\mathtt{LSSE}}=\hat{\bar{\mathbf{c}}}^{\mathtt{LSSE}}_{\mathtt{up}}$.  In Section~VI, Fig.~\ref{Fig:Seq_LSSE}, we  show numerically that even for short sequence lengths, this upper bound is not loose. 

Defining the estimation error as $\mathbf{e}^{\mathtt{LSSE}}_{\mathtt{up}} = \bar{\mathbf{c}} - \hat{\bar{\mathbf{c}}}^{\mathtt{LSSE}}_{\mathtt{up}}$, the expected square error norm is obtained as
\begin{IEEEeqnarray}{lll} \label{Eq:LSSE_ErrorNorm_Expected1}
 \mathbbmss{E}_{\mathbf{r},\bar{\mathbf{c}}}\left\{\|\mathbf{e}^{\mathtt{LSSE}}_{\mathtt{up}}\|^2\right\} \nonumber \\
 = \mathbbmss{E}_{\mathbf{r},\bar{\mathbf{c}}}\left\{\mathrm{tr}\left\{\left(\bar{\mathbf{c}} - \left(\mathbf{S}^T \mathbf{S}\right)^{-1}  \mathbf{S}^T \mathbf{r} \right)\left(\bar{\mathbf{c}} - \left(\mathbf{S}^T \mathbf{S}\right)^{-1}  \mathbf{S}^T \mathbf{r} \right)^T\right\} \right\} \nonumber \\
 =  \mathbbmss{E}_{\mathbf{r},\bar{\mathbf{c}}}\bigg\{ \mathrm{tr}\left\{ \left(\mathbf{S}^T \mathbf{S}\right)^{-1} \mathbf{S}^T \mathbf{r} \mathbf{r}^T \mathbf{S} \left(\mathbf{S}^T \mathbf{S}\right)^{-1} \right\} \nonumber \\
 \qquad\qquad - 2 \mathrm{tr}\left\{ \bar{\mathbf{c}}\mathbf{r}^T \mathbf{S} \left(\mathbf{S}^T \mathbf{S}\right)^{-1}  \right\} + \mathrm{tr}\left\{\bar{\mathbf{c}} \bar{\mathbf{c}}^T\right\} \bigg\}. 
\end{IEEEeqnarray}
 Next, we calculate the expectation with respect to $(\mathbf{r},\bar{\mathbf{c}})$  in (\ref{Eq:LSSE_ErrorNorm_Expected1})  in two steps, namely first with respect to $\mathbf{r}$ conditioned on $\bar{\mathbf{c}}$ and then with respect to $\bar{\mathbf{c}}$. Exploiting  $\mathbbmss{E}_{\mathbf{X}}\left\{\mathrm{tr}\left\{\mathbf{AXB}\right\}\right\} = \mathrm{tr}\left\{\mathbf{A}\mathbbmss{E}_{\mathbf{X}}\left\{\mathbf{X}\right\}\mathbf{B}\right\}$ and  $\mathbbmss{E}_{\mathbf{x}}\left\{\mathbf{x}\mathbf{x}^T\right\}  = \boldsymbol{\lambda}\boldsymbol{\lambda}^T + \mathrm{diag}\{\boldsymbol{\lambda}\}$,  $\mathbbmss{E}\left\{\|\mathbf{e}^{\mathtt{LSSE}}_{\mathtt{up}}\|^2\right\}$ can be calculated as
\begin{IEEEeqnarray}{lll} \label{Eq:LSSE_ErrorNorm_Expected2}
\mathbbmss{E}_{\bar{\mathbf{c}}}\mathbbmss{E}_{\mathbf{r}|\bar{\mathbf{c}}}\left\{\|\mathbf{e}^{\mathtt{LSSE}}_{\mathtt{up}}\|^2\right\}   \nonumber \\
=   \mathbbmss{E}_{\bar{\mathbf{c}}}\bigg\{ \mathrm{tr}\left\{ \left(\mathbf{S}^T \mathbf{S}\right)^{-1} \mathbf{S}^T \left(\mathbf{S}\bar{\mathbf{c}}\bar{\mathbf{c}}^T \mathbf{S}^T\right) \mathbf{S} \left(\mathbf{S}^T \mathbf{S}\right)^{-1} \right\} \nonumber \\
\qquad - 2 \mathrm{tr}\left\{ \bar{\mathbf{c}}\bar{\mathbf{c}}^T \mathbf{S}^T \mathbf{S} \left(\mathbf{S}^T \mathbf{S}\right)^{-1}  \right\} + \mathrm{tr}\left\{\bar{\mathbf{c}} \bar{\mathbf{c}}^T\right\} \nonumber \\
\qquad + \mathrm{tr}\left\{  \left(\mathbf{S}^T \mathbf{S}\right)^{-1} \mathbf{S}^T \mathrm{diag}\left\{ \mathbf{S}\bar{\mathbf{c}} \right\} \mathbf{S} \left(\mathbf{S}^T \mathbf{S}\right)^{-1} \right\} \bigg\} \qquad \nonumber \\
 =\mathbbmss{E}_{\bar{\mathbf{c}}}\left\{ \mathrm{tr}\left\{ \mathbf{S} \left(\mathbf{S}^T \mathbf{S}\right)^{-2} \mathbf{S}^T  \mathrm{diag}\left\{ \mathbf{S}\bar{\mathbf{c}} \right\} \right\} \right\} \nonumber \\
 = \mathrm{tr}\left\{ \mathbf{S}^T \mathrm{vdiag}\left\{\mathbf{S} \left(\mathbf{S}^T \mathbf{S}\right)^{-2} \mathbf{S}^T\right\}  \boldsymbol{\mu}_{\bar{\mathbf{c}}}^T  \right\}.
\end{IEEEeqnarray}

\begin{remk}   
From (\ref{Eq:LSSE_ErrorNorm_Expected2}), we observe that the square error norm of $\bar{\mathbf{c}}$ for the LSSE estimator can be factorized into two terms, a first term $\mathbf{S}^T \mathrm{vdiag}\left\{\mathbf{S} \left(\mathbf{S}^T \mathbf{S}\right)^{-2} \mathbf{S}^T\right\}$, which solely depends on the training sequence, multiplied by a second term $\boldsymbol{\mu}_{\bar{\mathbf{c}}}$, which depends on the MC channel. \hfill$\QED$

\end{remk}

The evaluation of the expression in (\ref{Eq:LSSE_ErrorNorm_Expected2}) can be numerically challenging due to the required inversion of matrix $\mathbf{S}^T \mathbf{S}$, especially when one of the eigen-values of $\mathbf{S}^T \mathbf{S}$ is close to zero. One way to cope with this problem is to eliminate all sequences resulting in   close-to-zero eigen-values for matrix  $\mathbf{S}^T \mathbf{S}$ during the search. Formally, we can adopt the following search criterion for LSSE-based training sequence design
\begin{IEEEeqnarray}{lll} \label{Eq:LSSE_Eig}
 \mathbf{s}^{\mathtt{LSSE}} = \underset{\mathbf{s} \in \widetilde{\mathcal{S}}  }{\mathrm{argmin}} \,\,  \mathrm{tr}\left\{ \mathbf{S}^T \mathrm{vdiag}\left\{\mathbf{S} \left(\mathbf{S}^T  \mathbf{S} \right)^{-2} \mathbf{S}^T\right\}  \boldsymbol{\mu}_{\bar{\mathbf{c}}}^T  \right\}, \quad\,\,
\end{IEEEeqnarray}
where $\widetilde{\mathcal{S}} =\left\{\mathbf{s}\in\mathcal{S}^K\big| |x|>  \varepsilon,\,\,\,\,\forall x\in\mathrm{eig}\left\{\mathbf{S}^T\mathbf{S}\right\}  \right\}$ and $\varepsilon$ is a small number which guarantees that the eigen-values of matrix $\mathbf{S}^T  \mathbf{S}$ are not close to zero, e.g., in Section~VI, we choose  $\varepsilon=10^{-9}$.

\subsection{LMMSE-Based Training Sequence Design}

Similar to the LSSE-based sequence design, the LMMSE estimate $\hat{\bar{\mathbf{c}}}^{\mathtt{MMSE}}$ in Lemma~\ref{Lem:MMSE} can be used as the basis for sequence design. In this paper,  we choose the optimal LMMSE-based training sequence such that the upper bound on the MMSE estimation error is minimized, i.e., $\mathbbmss{E}_{\mathbf{r},\bar{\mathbf{c}}}\left\{\|\mathbf{e}^{\mathtt{MMSE}}_{\mathtt{up}}\|^2\right\}$ in (\ref{Eq:MMSE_ErrorNorm}). In particular, substituting $\mathbf{F}^{\mathtt{MMSE}}$ in (\ref{Eq:MMSE_Sol}) into (\ref{Eq:MMSE_ErrorNorm}), simplifying the resulting expression for $\mathbbmss{E}_{\mathbf{r},\bar{\mathbf{c}}}\left\{\|\mathbf{e}^{\mathtt{MMSE}}_{\mathtt{up}}\|^2\right\}$, and removing the terms which do not depend on the sequence, we obtain the following criterion for  the optimal LMMSE-based training sequence
\begin{IEEEeqnarray}{lll} \label{Eq:MMSE_Seq}
 \mathbf{s}^{\mathtt{MMSE}} = \underset{\forall\mathbf{s}\in\mathcal{S}^K  }{\mathrm{argmax}} \,\,  \mathrm{tr}\left\{ \boldsymbol{\Phi}_{\bar{\mathbf{c}}\bar{\mathbf{c}}}^T \mathbf{S}^T  \left(\mathbf{S}\boldsymbol{\Phi}_{\bar{\mathbf{c}}\bar{\mathbf{c}}} \mathbf{S}^T +  \mathrm{diag}\left\{\mathbf{S}\boldsymbol{\mu}_{\bar{\mathbf{c}}}\right\} \right)^{-1}
 \mathbf{S} \boldsymbol{\Phi}_{\bar{\mathbf{c}}\bar{\mathbf{c}}} \right\}.  \quad
\end{IEEEeqnarray}
We note that from (\ref{Eq:MMSE_Qvar}a), we can conclude that matrix $\mathbf{S}\boldsymbol{\Phi}_{\bar{\mathbf{c}}\bar{\mathbf{c}}} \mathbf{S}^T +  \mathrm{diag}\left\{\mathbf{S}\boldsymbol{\mu}_{\bar{\mathbf{c}}}\right\}$  is a positive definite matrix, i.e., all eigen-values are positive.

\begin{remk}
We base our training sequence designs on the LSSE and LMMSE estimators (and not the ML and MAP estimators) because these two estimators lend themselves to elegant closed-form solutions for the estimated CIR which leads to relatively simpler criteria for training sequence design, cf. (\ref{Eq:LSSE_Eig}) and (\ref{Eq:MMSE_Seq}). In Section~VI, we employ an exhaustive computer-based search to find the optimal LSSE-based and LMMSE-based training sequences by evaluating  (\ref{Eq:LSSE_Eig}) and (\ref{Eq:MMSE_Seq}), respectively. We note that the complexity of the exhaustive search can be accommodated since the optimal training sequence is obtained offline and used for online CIR estimation.  Nevertheless, to reduce the complexity of the exhaustive search, one can develop systematic approaches to solve the optimization problems in (\ref{Eq:LSSE_Eig}) and (\ref{Eq:MMSE_Seq}). This constitutes an interesting research problem which is beyond the scope of this paper and left for future work. \hfill$\QED$
\end{remk}

\subsection{ISI-Free Training Sequence Design}

One simple approach to estimate the CIR is to construct a training sequence such that ISI is avoided during estimation. In this case, in each symbol interval, the receiver will observe molecules  which have been released by the transmitter in only one symbol interval and not in multiple symbol intervals.  To this end, the transmitter releases $N^{\mathtt{Tx}}$  molecules every $L+1$ symbol intervals and remains silent for the rest of the symbol intervals. This corresponds to the ON-OFF keying, i.e., $s[k]\in\{0,1\}$, which is a special case of the general CSK modulation adopted in this paper.  In particular,  the sequence $\mathbf{s}$ is constructed as follows:
\begin{IEEEeqnarray}{lll} \label{Eq:Seq_ISIfree}
 s[k]=\begin{cases}
 1,\quad &\mathrm{if} \,\,\frac{k-k_0}{L+1}\in \mathbb{Z}\\
 0, & \mathrm{otherwise}
 \end{cases}
\end{IEEEeqnarray}
where $k\in\{1,\dots,K\}$, and $k_0$ is the index of the first symbol interval in which the transmitter releases molecules.   Moreover, for this training sequence, the CIR  can be straightforwardly estimated as
\begin{IEEEeqnarray}{lll} \label{Eq:CIR_ISIfree}
\hat{\bar{c}}^{\mathtt{ISIF}}_l = \frac{1}{|\mathcal{K}_l|} \Big[\sum_{k\in\mathcal{K}_l} \big[ r[k] - \hat{\bar{c}}^{\mathtt{ISIF}}_{\mathtt{n}} \big] \Big]^+, \quad l =1,\dots,L \,\,\, \IEEEyesnumber\IEEEyessubnumber \\
\hat{\bar{c}}^{\mathtt{ISIF}}_{\mathtt{n}} = \frac{1}{|\mathcal{K}_{\mathtt{n}}|} \sum_{k\in\mathcal{K}_{\mathtt{n}}} r[k],     \IEEEyessubnumber
\end{IEEEeqnarray}
where $\mathcal{K}_l=\big\{k|\frac{k-k_0-l+1}{L+1}\in\mathbb{Z}\,\,\wedge\,\,k\in\{1,\dots,K\} \big\}$, $\mathcal{K}_{\mathtt{n}} = \big\{k|\frac{k-k_0-L}{L+1}\in\mathbb{Z} \,\,\wedge\,\,k\in\{1,\dots,K\}  \big\}$, and $[\cdot]^+$ is needed to ensure that all estimated channel taps are non-negative, i.e., $\hat{\bar{\mathbf{c}}}^{\mathtt{ISIF}}\geq\mathbf{0}$ holds.

\begin{remk}
The ISI-free training sequence in (\ref{Eq:Seq_ISIfree}) permits the use of the simple estimator in (\ref{Eq:CIR_ISIfree}). Thereby,  the mathematical calculation for $\hat{\bar{c}}^{\mathtt{ISIF}}_{\mathtt{n}}$  boils down to averaging over those elements of the observation vector $\mathbf{r}$ which are only affected by the external noise. Similarly, the computation of $\hat{\bar{c}}^{\mathtt{ISIF}}_l$ basically reduces to averaging over those elements of $\mathbf{r}$ which are affected by $\bar{c}_l$ and the impact of the estimated mean of the external noise $\hat{\bar{c}}^{\mathtt{ISIF}}_{\mathtt{n}}$ is removed. The ISI-free training sequence in (\ref{Eq:Seq_ISIfree}) and the corresponding simple estimator in (\ref{Eq:CIR_ISIfree}) are suitable options for applications where nanoreceivers with low computational capability have to be employed for CIR estimation.
We note that, for the ISI-free training sequence in (\ref{Eq:Seq_ISIfree}), the asymptotic solutions of the ML, LSSE, MAP, and LMMSE estimates as $K\to\infty$ reduce to the estimate given in (\ref{Eq:CIR_ISIfree}), i.e., $\hat{\bar{\mathbf{c}}}^{\mathtt{ML}}=\hat{\bar{\mathbf{c}}}^{\mathtt{LSSE}}=\hat{\bar{\mathbf{c}}}^{\mathtt{MAP}}=\hat{\bar{\mathbf{c}}}^{\mathtt{MMSE}}=\hat{\bar{\mathbf{c}}}^{\mathtt{ISIF}}$.
\hfill$\QED$
\end{remk}

\section{Performance Evaluation}

\begin{figure*}[!tbp]
  \centering
  \begin{minipage}[b]{0.47\textwidth}
  \centering
\resizebox{1.05\linewidth}{!}{\psfragfig{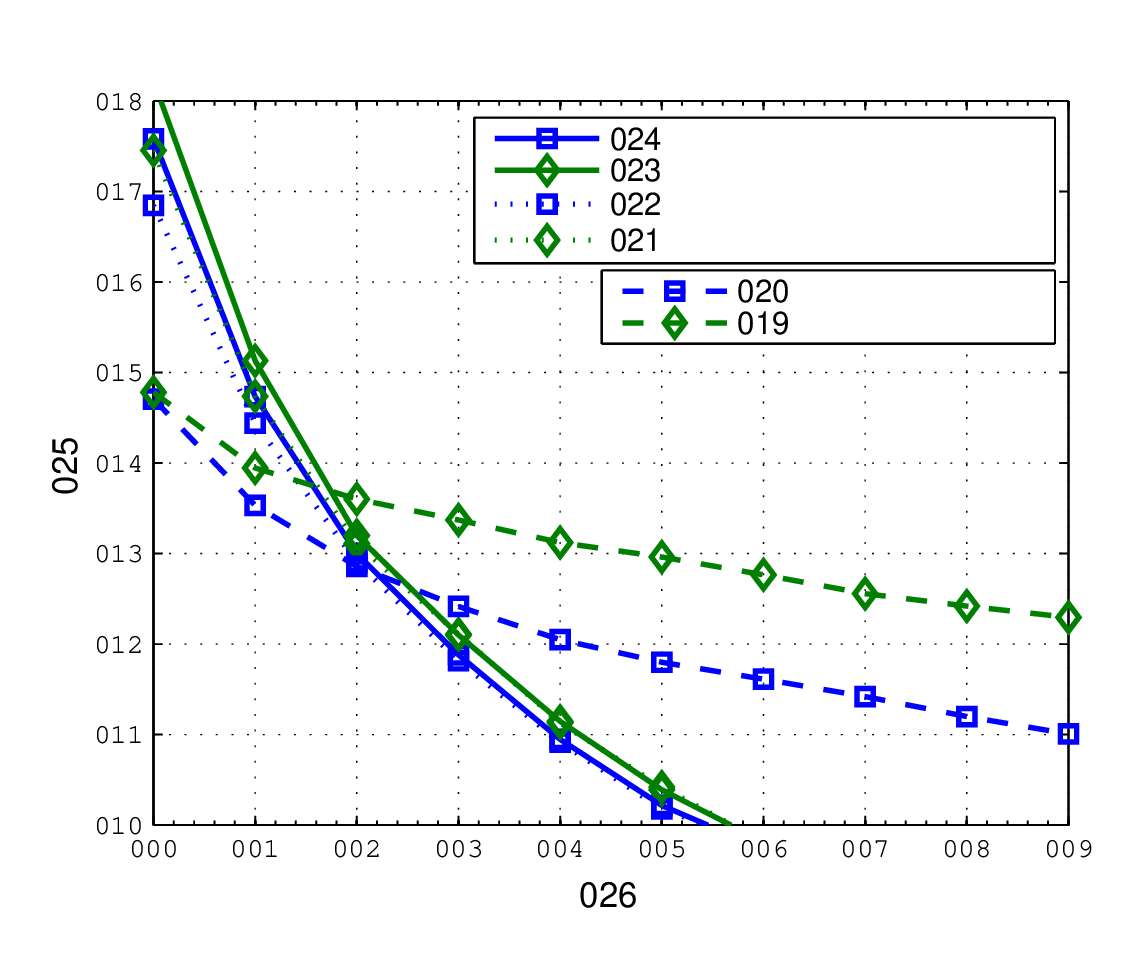}}  
\caption{Norm of the estimation error mean, $\left\|\mathbbmss{E}\left\{ \mathbf{e} \right\} \right\|$, in dB vs. the training sequence length, $K$, for $L=3$ and $\sigma^2=0.1$. }
\label{Fig:Mean_K}
  \end{minipage}
  \hfill
  \begin{minipage}[b]{0.1\textwidth}
  \end{minipage}
  \hfill
    \begin{minipage}[b]{0.47\textwidth}
  \centering
\resizebox{1.05\linewidth}{!}{\psfragfig{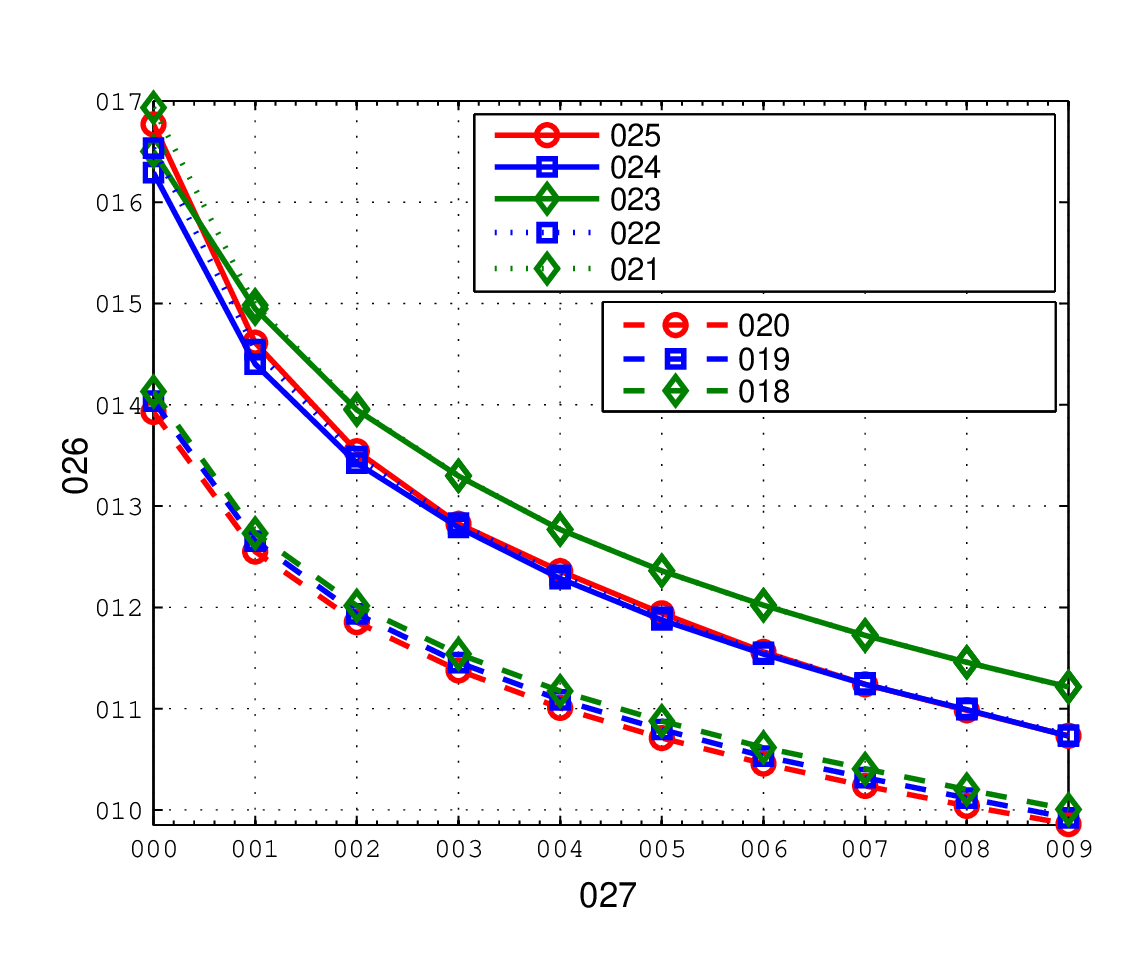}} 
\caption{Estimation error variance, $\mathbbmss{E}\left\{\left\| \mathbf{e} - \mathbbmss{E}\left\{ \mathbf{e} \right\} \right\|^2\right\} $, in dB vs. the training sequence length, $K$, for $L=3$ and $\sigma^2=0.1$. }
\label{Fig:Var_K}
  \end{minipage}
    \hfill
  \begin{minipage}[b]{0.02\textwidth}
  \end{minipage}\vspace{-0.4cm}
\end{figure*}

In this section, we evaluate the performances of the different estimation techniques and training sequence designs developed in this paper for ON-OFF keying signaling, i.e., $s[k]\in\{0,1\}$. For simplicity, for the results provided in this section, we generate  CIR $\bar{\mathbf{c}}$ based on (\ref{Eq:Cons_CIR}) and (\ref{Eq:Consentration}). However, we emphasize that the proposed estimation framework is not limited to the particular  channel and receiver models assumed in (\ref{Eq:Cons_CIR}) and (\ref{Eq:Consentration}). We use (\ref{Eq:Cons_CIR}) and (\ref{Eq:Consentration}) only to obtain a CIR $\bar{\mathbf{c}}$ which is representative of a typical CIR in MC.  In particular, we assume a point source with impulsive molecule release and $N^{\mathtt{Tx}}=10^5$, a fully transparent spherical receiver with radius $50$ nm,  and an unbounded environment with $D=4.365\times10^{-10} \,\frac{\text{m}^2}{\text{s}}$ \cite{Arman_AF}. The distance between the transmitter and the receiver is assumed to be $|\mathbf{a}| = 400$ nm. The receiver counts the number of molecules once per symbol interval at time $T_{\mathrm{smp}} = {\mathrm{argmax}}_{\,t}\,\,\bar{\mathcal{C}}(\mathbf{a},t)$ after the beginning of the symbol interval where $\bar{\mathcal{C}}(\mathbf{a},t)$ is computed based on (\ref{Eq:Consentration}). The noise mean is chosen as $\bar{c}_{\mathtt{n}} = 0.2 {\mathrm{max}}_{\,t}\,\,\bar{\mathcal{C}}(\mathbf{a},t)$. Furthermore, the symbol duration and the number of taps $L$ are chosen such that $\bar{c}_{L+1}<0.05 \bar{c}_1$. Based on the aforementioned assumptions, we obtain a typical/default value for the CIR vector which is denoted by $\bar{\mathbf{c}}^{\mathtt{def}}$, e.g., $\bar{\mathbf{c}}^{\mathtt{def}}=[60.22,\,\,   11.76,  \,\,  5.13, \,\,  12.04]^T$ for $L=3$. In practice, however,  there may be random variations in the underlying channel parameters, e.g., $D$, $|\mathbf{a}|$,  $N^{\mathtt{Tx}}$, which leads to random variations in the CIR. To take this into account, we assume that the CIR vector in each estimation block is a realization of a RV  according to $\bar{\mathbf{c}}=\left[\bar{\mathbf{c}}^{\mathtt{Gaus}}\right]^+=\left[\bar{\mathbf{c}}^{\mathtt{def}}+\sigma\mathrm{diag}\{\bar{\mathbf{c}}^{\mathtt{def}}\}\mathcal{N}\left(\mathbf{0},\mathbf{I}\right)\right]^+$, where $\sigma$ is the standard deviation of RV $\bar{\mathbf{c}}^{\mathtt{Gaus}}$, see Remark~\ref{Remk:Fc} for further discussion.  We note that as $\sigma\to 0$, the MC channel becomes deterministic,  and for large $\sigma$,  the MC channel is highly stochastic.

\begin{figure*}[!tbp]
  \centering
  \begin{minipage}[b]{0.47\textwidth}
  \centering
\resizebox{1.05\linewidth}{!}{\psfragfig{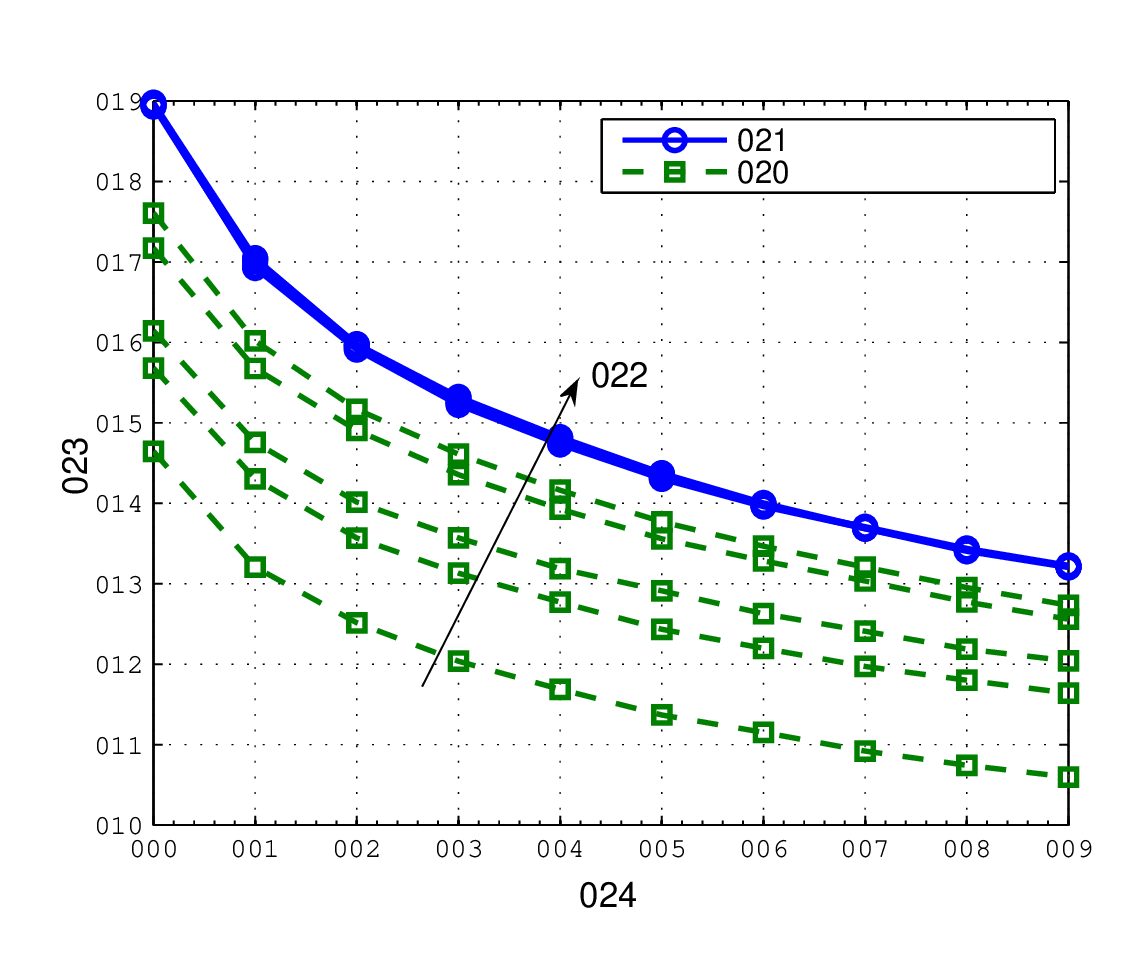}} 
\caption{Estimation error variance, $\mathbbmss{E}\left\{\left\| \mathbf{e} - \mathbbmss{E}\left\{ \mathbf{e} \right\} \right\|^2\right\} $, in dB vs. the training sequence length, $K$, for $L=3$ and $\sigma^2=[0.01,0.05,0.1,0.5,1]$. }
\label{Fig:Var_K_CIR}
  \end{minipage}
    \hfill
  \begin{minipage}[b]{0.1\textwidth}
  \end{minipage}
  \hfill
  \begin{minipage}[b]{0.47\textwidth}
  \centering
\resizebox{1.05\linewidth}{!}{\psfragfig{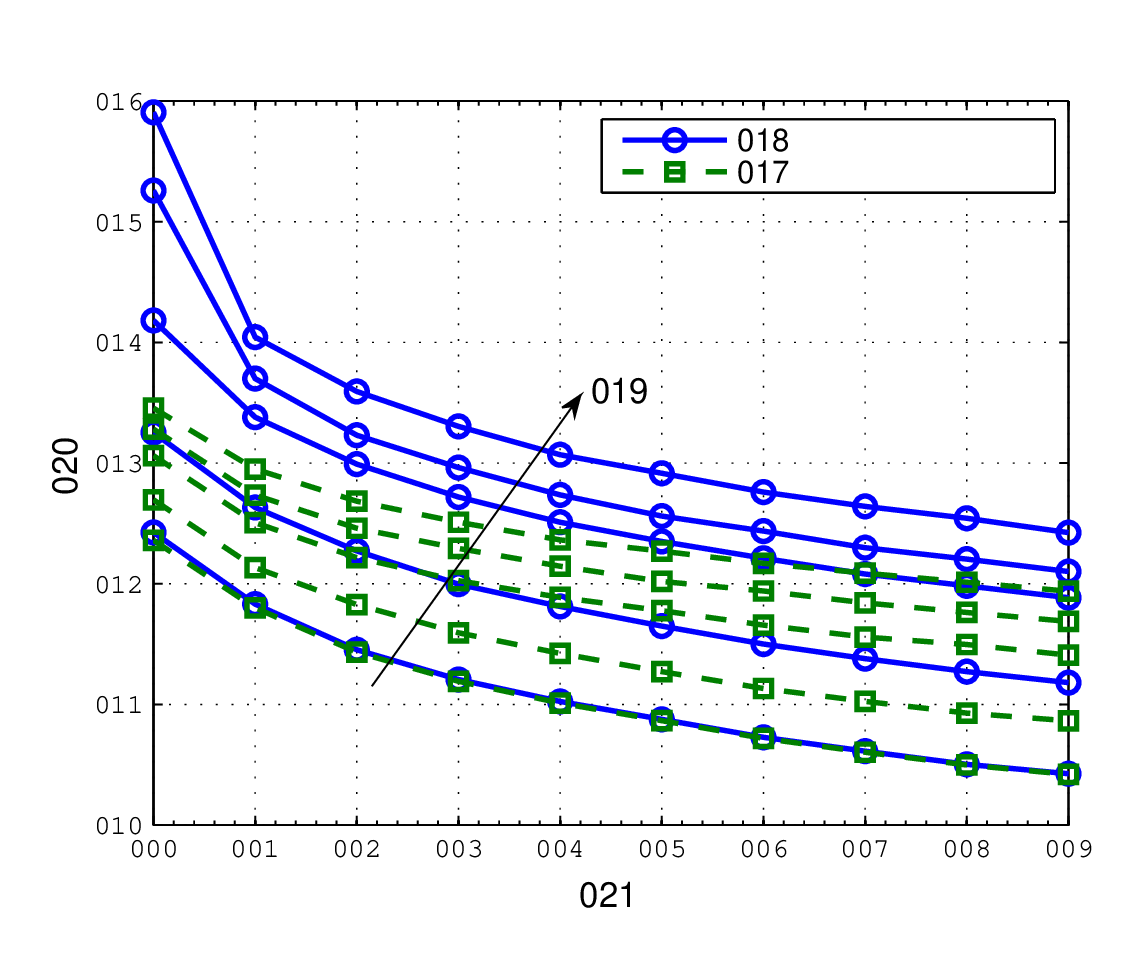}} 
\caption{Normalized estimation error variance, $\overline{\mathrm{Var}}_{\mathbf{e}}$, in dB vs. the training sequence length, $K$, for $L\in\{1,2,3,4,5\}$ and $\sigma^2=0.1$. }
\label{Fig:Var_K_L}
  \end{minipage}
    \hfill
  \begin{minipage}[b]{0.02\textwidth}
  \end{minipage}\vspace{-0.4cm}
\end{figure*}

For the results reported in Figs.~\ref{Fig:Mean_K}-\ref{Fig:Var_K_L}, the training sequences  are constructed by concatenating $n$ copies of the binary sequence $[1 1 0 0 1 0 0 1 0 1]$ of length $10$, i.e., $K=10n$. Moreover, the results reported in Figs.~\ref{Fig:Mean_K}-\ref{Fig:Var_K_L} are Monte Carlo simulations where each point of the curves is obtained by averaging over $10^5$ random realizations of observation vector $\mathbf{r}$ and $\bar{\mathbf{c}}$. In the following, in Figs.~\ref{Fig:Mean_K} and \ref{Fig:Var_K}, we present results for $L=3$ and $\sigma^2=0.1$.  In particular, in Fig.~\ref{Fig:Mean_K}, we show the norm of the mean of the estimation error, $\left\|\mathbbmss{E}\left\{ \mathbf{e} \right\} \right\|$, where $\mathbf{e} = \hat{\bar{\mathbf{c}}}-\bar{\mathbf{c}}$, in dB vs. the training sequence length, $K$. Since for a given $\bar{\mathbf{c}}$,  $\mathbbmss{E}_{\mathbf{r}|\bar{\mathbf{c}}}\left\{ \mathbf{e} \right\} = \mathbbmss{E}_{\mathbf{r}|\bar{\mathbf{c}}}\left\{ \hat{\bar{\mathbf{c}}} \right\} -  \bar{\mathbf{c}} $ holds, $\left\|\mathbbmss{E}\left\{ \mathbf{e} \right\} \right\|=\left\|\mathbbmss{E}_{\bar{\mathbf{c}}}\mathbbmss{E}_{\mathbf{r}|\bar{\mathbf{c}}}\left\{ \mathbf{e} \right\} \right\|$ is a measure for the \textit{average} bias of the estimate $\hat{\bar{\mathbf{c}}}$. We observe from Fig.~\ref{Fig:Mean_K} that the mean of the estimation error decreases as the sequence length increases. Therefore, the proposed estimators are biased for short sequence lengths, but as the sequence length increases, they become asymptotically unbiased, i.e., $\mathbbmss{E}\left\{\hat{\bar{\mathbf{c}}}\right\} \to \bar{\mathbf{c}}$ as $K\to\infty$, cf. Corollary~\ref{Corol:LSSE_Unbiased} and Remarks~\ref{Remk:ML_Unbiased}, \ref{Remk:ML_LSSE_AsymUnbiased}, and \ref{Remk:MAP_MMSE_Unbiased}. 

In Fig.~\ref{Fig:Var_K}, we show the estimation error variance\footnote{We note that the  estimation error variance \textit{normalized} to $\|\mathbbmss{E}\left\{\bar{\mathbf{c}}\right\}\|^2$ yields very small values. For instance, in Fig.~\ref{Fig:Var_K},  the value of the normalized estimation error variance  is approximately $36$~dB less than the value of the estimation error variance.}, $\mathbbmss{E}\left\{\left\| \mathbf{e} - \mathbbmss{E}\left\{ \mathbf{e} \right\} \right\|^2\right\} $, in dB vs. the training sequence length, $K$.  As expected, the variance of the estimation error decreases with increasing  training sequence length. For CIR estimation without statistical channel knowledge, we observe in Fig.~\ref{Fig:Var_K} that the gap between the error variances of the suboptimal ML-based (LSSE-based) estimate and  the optimal ML (LSSE) estimate is small for short sequence lengths and vanishes  as $K\to\infty$. Furthermore, for the considered set of parameters, the ML estimator outperforms the LSSE estimator by a margin of less than $1$~dB. These  results  suggest  that  the  simple  LSSE  estimator  provides a favorable complexity-performance tradeoff for CIR estimation in  the  considered  MC  system. For CIR estimation with statistical channel knowledge, we observe from Fig.~\ref{Fig:Var_K} that the error variance of the LMMSE estimate is slightly higher than that of the MAP estimate which reveals the effectiveness of the proposed LMMSE estimator. Additionally, for short sequence lengths,  a reduction of up to $5$~dB  in the estimation error variance is attainable with the MAP/LMMSE estimators as compared to the ML/LSSE estimators at the cost of the required acquisition of statistical channel  knowledge. However, this gain vanishes as $K\to\infty$. Finally, we note that for short sequence lengths, the variances of the proposed estimators can even be  lower than the corresponding classical/Bayesian CR bound as these estimators are biased and the  CR bound is a valid lower bound only for unbiased estimators, see Fig.~\ref{Fig:Mean_K}. However, as $K$ increases, all estimators become asymptotically  unbiased, see Fig.~\ref{Fig:Mean_K}. Fig.~\ref{Fig:Var_K} shows that, for large $K$, the error variance of the ML estimator coincides with the classical CR bound and the error variance of the MAP estimator is very close to the Bayesian CR bound.

To further study the performance of the proposed estimators, in Figs.~\ref{Fig:Var_K_CIR} and \ref{Fig:Var_K_L}, we show results only for the suboptimal LSSE estimator and the LMMSE estimator. Note that these linear estimators are easier to implement compared to the other proposed estimators. In Fig.~\ref{Fig:Var_K_CIR}, we show the estimation error variance, $\mathbbmss{E}\left\{\left\| \mathbf{e} - \mathbbmss{E}\left\{ \mathbf{e} \right\} \right\|^2\right\} $, in dB vs. the training sequence length, $K$  for $L=3$ and $\sigma^2\in\{1,0.5,0.1,0.05,0.01\}$. As expected, the error variance of the LMMSE estimator decreases as $\sigma^2$ decreases. In particular, when $\sigma^2\to\infty$, the statistical channel knowledge becomes useless and when $\sigma^2\to 0$, the first order statistics $\boldsymbol{\mu}_{\bar{\mathbf{c}}}$ become a perfectly precise estimate of $\bar{\mathbf{c}}$. Thereby, depending on the values of $K$ and $\sigma^2$, we observe from Fig.~\ref{Fig:Var_K_CIR} that the LMMSE estimator achieves gains of $1$-$9$ dB compared to the LSSE estimator at the expense of requiring knowledge of the first and second order statistics of the CIR.   Moreover, as suggested by (\ref{Eq:LSSE_ErrorNorm_Expected2}), the LSSE error variance depends only on $\boldsymbol{\mu}_{\bar{\mathbf{c}}}$ which is approximately constant for all  curves in Fig.~\ref{Fig:Var_K_CIR}. In particular, the small change in the LSSE error variance in Fig.~\ref{Fig:Var_K_CIR} is due to the fact that although by increasing $\sigma^2$, the mean of $\bar{\mathbf{c}}^{\mathtt{Gaus}}$ remains unchanged, the mean of  $\bar{\mathbf{c}}=\left[\bar{\mathbf{c}}^{\mathtt{Gaus}}\right]^+$ changes since the instances, in which the elements of $\bar{\mathbf{c}}^{\mathtt{Gaus}}$ are negative, occur more frequently.

In order to compare the performances of the considered estimators for different numbers of channel memory taps, $L$, we define the normalized variance of the estimation error  as 
\begin{IEEEeqnarray}{lll} \label{Eq:Var_Norm}
 \overline{\mathrm{Var}}_{\mathbf{e}} \,\,&=  \frac{\mathbbmss{E}\left\{\|\mathbf{e}\|^2\right\} - \|\mathbbmss{E}\left\{\mathbf{e}\right\}\|^2 }{\|\mathbbmss{E}\left\{\bar{\mathbf{c}}\right\}\|^2}. 
\end{IEEEeqnarray}
In Fig.~\ref{Fig:Var_K_L}, we show the normalized estimation error variance, $\overline{\mathrm{Var}}_{\mathbf{e}}$, in dB vs. the training sequence length, $K$, for $L\in\{1,2,3,4,5\}$ and $\sigma^2=0.1$. Thereby, we observe that the error variance increases as the number of memory taps increases. This is due to the fact that  as $L$ increases,  the number of unkown parameters which have to be estimated increases. Furthermore, Fig.~\ref{Fig:Var_K_L} shows that the gap between the error variances of the LMMSE and LSSE estimators increases as $L$ increases.

Next, we investigate the performances of the optimal LSSE-/LMMSE-based and ISI-free training sequence designs developed in Section~V. Here, we employ a computer-based search to find the optimal sequences  based on the LSSE-based criterion in (\ref{Eq:LSSE_Eig}) where $\varepsilon=10^{-9}$ and the LMMSE-based criterion in (\ref{Eq:MMSE_Seq}). We consider short sequence lengths, i.e., $K\leq 20$, due to the exponential increase of the computational complexity of the exhaustive search with respect to the sequence length. Moreover, since there are $L+1$ unknown parameters, we require at least $L+1$ observations for estimation, i.e., $K-L+1\geq L+1$ or equivalently $K\geq 2L$. In Table~I, we present the optimal sequences obtained for  $L\in\{1,3,5\}$, $K\in\{10,20\}$, and $\sigma^2=0.1$.  We note that the optimal sequences which are obtained from  (\ref{Eq:LSSE_Eig}) and (\ref{Eq:MMSE_Seq}) are not unique and only one of the optimal sequences is shown in Table~I for each value of $K$ and $L$.  The optimal sequence shown in red font in Table~I is an ISI-free sequence as defined in (\ref{Eq:Seq_ISIfree}). 

\begin{table*}
\label{Table:OptSeq}
\caption{Examples of Optimal LSSE/LMMSE Sequences  Obtained by a Computer-Based Search \newline  for  $L\in\{1,3,5\}$, $K\in\{10,20\}$, and $\sigma^2=0.1$.} 
\begin{center}
\scalebox{0.9}{
\begin{tabular}{|| c | l | c | c ||}
  \hline                  
 &   Criterion  &  $K=10$ &  $K=20$   \\ \hline
\multirow{2}{*}{ $L=1$ } & LSSE-based &   $\mathbf{s}^* = [1001101110]^T$ & $\mathbf{s}^* = [00101110111010101111]^T$ \\ 
& LMMSE-based & $\mathbf{s}^* = [1111000111]^T$ & $\mathbf{s}^* = [11011110101100010111]^T$\\ \hline
\multirow{2}{*}{ $L=3$ } & LSSE-based&   $\mathbf{s}^* = [1100001001]^T$ & 
 $\mathbf{s}^* = [10101101101101100000]^T$ \\ 
 & LMMSE-based & $\mathbf{s}^* = [0001101011]^T$  & $\mathbf{s}^* = [00110101100000111011]^T$ \\ \hline
\multirow{2}{*}{ $L=5$ } & LSSE-based&  {\color[rgb]{1,0,0} $\,\,\mathbf{s}^* = [1000001000]^T$ } & 
 $\mathbf{s}^* = [11111110000100001000]^T$ \\ 
 & LMMSE-based & $\mathbf{s}^* = [0000010110]^T$  & $\mathbf{s}^* = [00000011001010100111]^T$ \\ \hline
\end{tabular}
}
\end{center}
 
\end{table*}

\begin{figure*}[!tbp]
  \centering
  \begin{minipage}[b]{0.47\textwidth}
  \centering
\resizebox{1.05\linewidth}{!}{\psfragfig{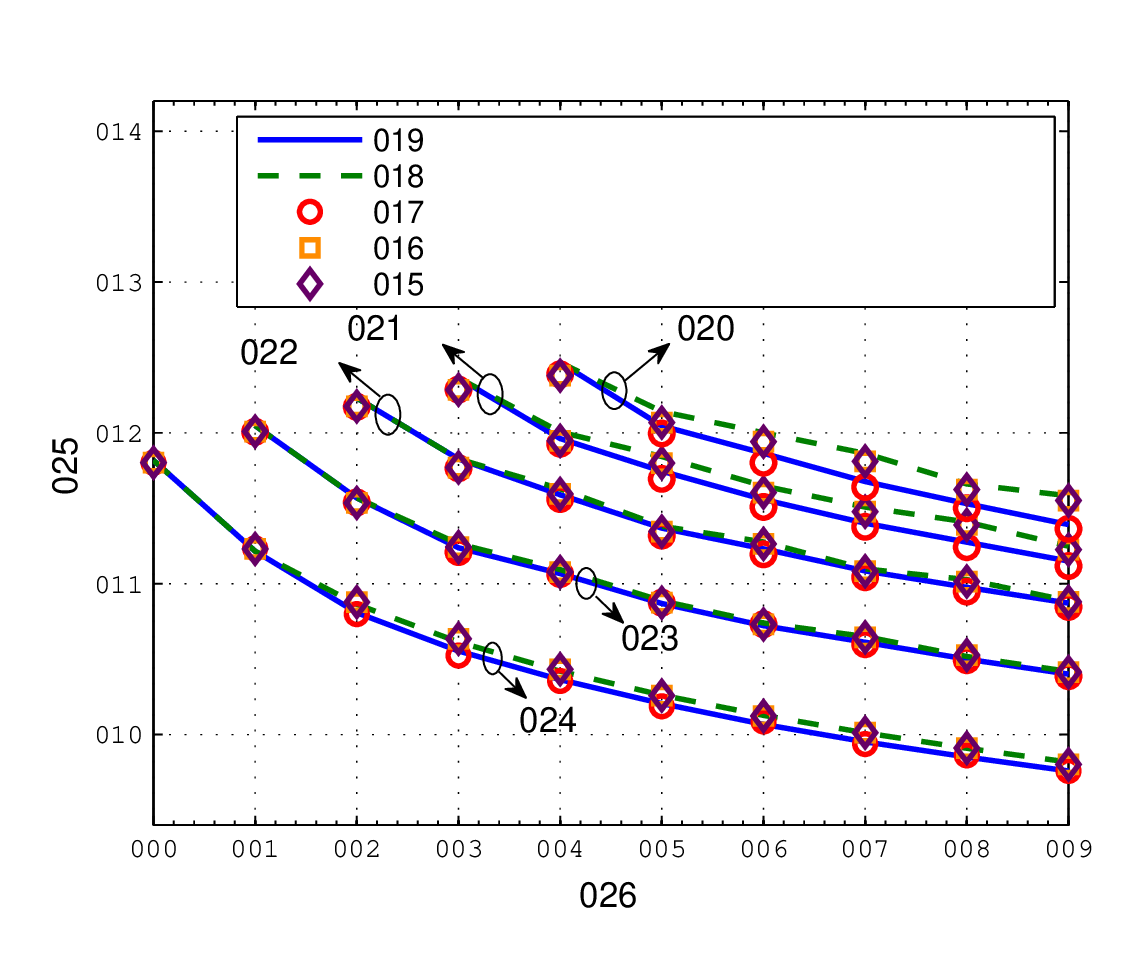}} 
\caption{Normalized LSSE estimation error variance, $\overline{\mathrm{Var}}_{\mathbf{e}}$, in dB vs. the training sequence length, $K$, for $L\in\{1,2,3,4,5\}$ and $\sigma^2=0.1$.  }
\label{Fig:Seq_LSSE}
  \end{minipage}
    \hfill
  \begin{minipage}[b]{0.1\textwidth}
  \end{minipage}
  \hfill
  \begin{minipage}[b]{0.47\textwidth}
  \centering
\resizebox{1.05\linewidth}{!}{\psfragfig{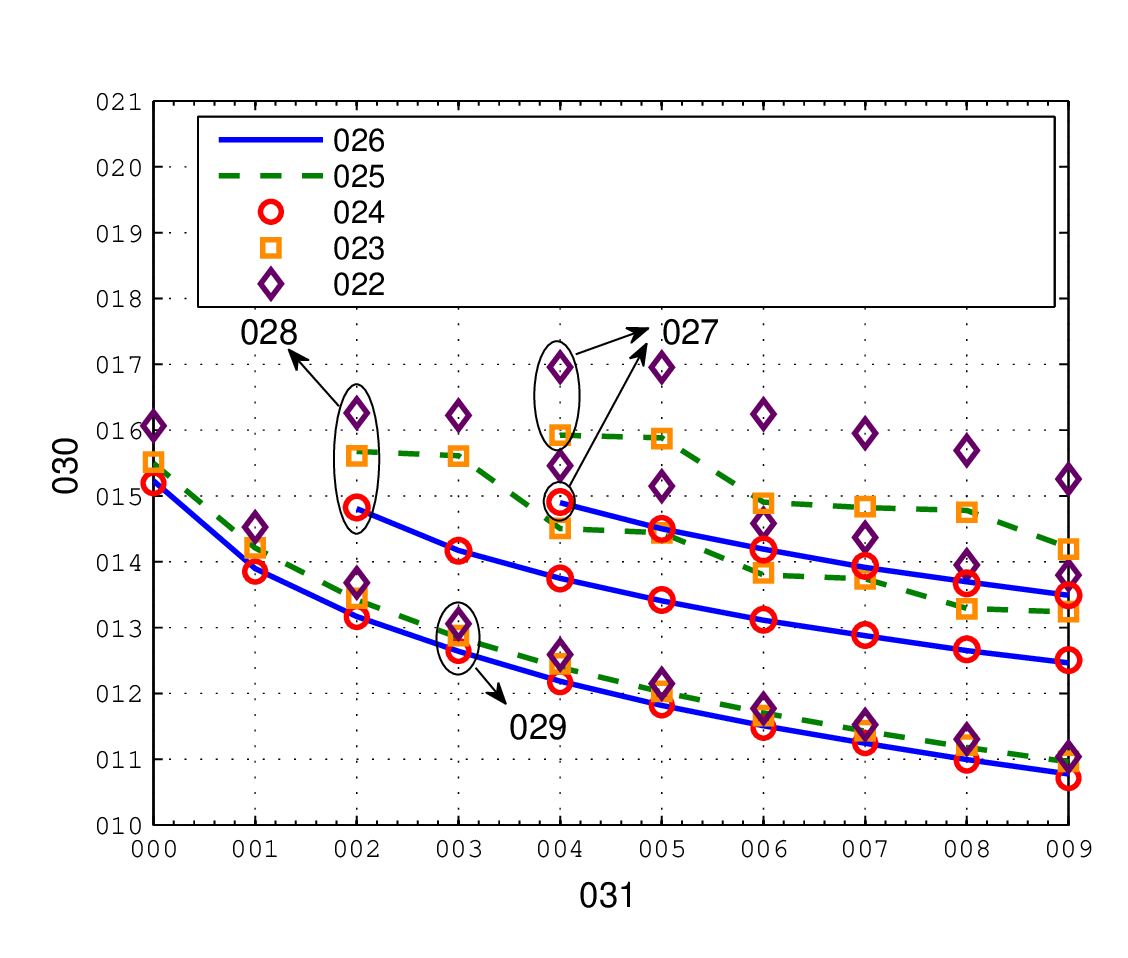}} 
\caption{Normalized LMMSE estimation error variance, $\overline{\mathrm{Var}}_{\mathbf{e}}$, in dB vs. the training sequence length, $K$, for $L\in\{1,3,5\}$ and $\sigma^2=0.1$.  }
\label{Fig:Seq_MMSE}
  \end{minipage}
    \hfill
  \begin{minipage}[b]{0.02\textwidth}
  \end{minipage}\vspace{-0.4cm}
\end{figure*}

In Fig.~\ref{Fig:Seq_LSSE}, we show the normalized LSSE estimation error variance, $\overline{\mathrm{Var}}_{\mathbf{e}}$, in dB vs. the training sequence length, $K$, for $L\in\{1,2,3,4,5\}$ and $\sigma^2=0.1$. Thereby, we report the analytical results for the upper bound in (\ref{Eq:LSSE_ErrorNorm_Expected2}) and  Monte Carlo simulation results for $10^5$ random realizations of $\mathbf{r}$ and $\bar{\mathbf{c}}$. For the ISI-free sequence, we also report the performance of the simple estimation method in (\ref{Eq:CIR_ISIfree}) in addition to the performance of LSSE estimation. Fig.~\ref{Fig:Seq_LSSE} confirms that  (\ref{Eq:LSSE_ErrorNorm_Expected2}) is a tight upper bound even for short sequence lengths. Moreover, we observe from Fig.~\ref{Fig:Seq_LSSE} that the difference between the error variances of the ISI-free sequence and the optimal sequence  increases as $L$ increases. This result suggests that for MC channels with small numbers of taps, a simple ISI-free training sequence is a suitable option.  Furthermore, as expected, the estimation error decreases with increasing  training sequence length and increases with increasing number of CIR coefficients.
 
In Fig.~\ref{Fig:Seq_MMSE}, we show the normalized LMMSE estimation error, $\overline{\mathrm{Var}}_{\mathbf{e}}$, in dB vs. the training sequence length, $K$, for $L\in\{1,3,5\}$ and $\sigma^2=0.1$. In general, we can observe a similar behavior as in Fig.~\ref{Fig:Seq_LSSE}, hence, we highlight only the differences. In particular, the simulation results for LMMSE estimation and the corresponding upper bound in (\ref{Eq:MMSE_ErrorNorm}) coincide which reveals that the effect of the instances for which the elements of  $\mathbf{F}^{\mathtt{MMSE}}\mathbf{r}$ are negative is negligible and hence $\hat{\bar{\mathbf{c}}}^{\mathtt{MMSE}} = \left[\mathbf{F}^{\mathtt{MMSE}}\mathbf{r}\right]^+\approx \mathbf{F}^{\mathtt{MMSE}}\mathbf{r}$. Moreover, for $L=3$ and $L=5$, there is a considerable gap between the variances of the MMSE estimators for the optimal LMMSE-based training sequence and the ISI-free sequence. 

 
\section{Conclusions}

In this paper, we developed a training-based CIR estimation framework which enables the acquisition of the CIR based on the observed number of molecules at the receiver due to emission of a  sequence of known numbers of molecules by the transmitter.  In particular, for the case when statistical channel knowledge is not available, we derived the optimal ML estimator, the suboptimal LSSE  estimator, and the classical CR lower bound. On the other hand, for the case when knowledge of the channel statistics is available, we provided the optimal MAP estimator, the suboptimal LMMSE estimator, and the Bayesian CR lower bound. Furthermore, we studied optimal LSSE-/LMMSE-based and suboptimal training sequence designs for the considered MC system. Our simulation results revealed that the simple LSSE and LMMSE estimators provide favorable tradeoffs between complexity and performance. Additionally, for LSSE estimation, the proposed simple ISI-free sequences achieve a similar performance as the optimal sequences whereas, for LMMSE estimation, the performance gap between the ISI-free and the optimal sequences is not negligible  specially for larger numbers of channel memory taps. 

In this paper, we investigated the performance of the proposed CIR estimators and training sequences in terms of the estimation error variance. Studying the effect of  the quality of the CIR estimates on the end-to-end bit error rate performance of MC systems is an interesting and important direction for future research.

\appendices

\section{Proof of Lemma~\ref{Lem:ML}}\label{App:ML}

The problem in (\ref{Eq:ML_Log}) is a convex optimization problem in variable $\bar{\mathbf{c}}$ because $g(\bar{\mathbf{c}})$ is a concave function in $\bar{\mathbf{c}}$ and the feasible set $\bar{\mathbf{c}}\geq \mathbf{0}$ is linear in $\bar{\mathbf{c}}$.  In particular, $\mathrm{ln}\left(\bar{\mathbf{c}}^T\mathbf{s}_k  \right)$ is concave  since $\bar{\mathbf{c}}^T\mathbf{s}_k$ is affine and the log-function is concave \cite[Chapter~3]{Boyd}. Therefore,  $g(\bar{\mathbf{c}})$ is a sum of weighted concave terms $r[k]\mathrm{ln}\left(\bar{\mathbf{c}}^T\mathbf{s}_k  \right)$ and affine terms $\bar{\mathbf{c}}^T\mathbf{s}_k$ which in turn yields a concave function \cite[Chapter~3]{Boyd}. For the constrained convex problem in (\ref{Eq:ML_Log}), the optimal solution falls into one of the following two categories:

\textit{Stationary Point:} In this case, the optimal solution is found by taking the derivative of  $g(\bar{\mathbf{c}})$ with respect to $\bar{\mathbf{c}}$ which is given in (\ref{Eq:ML_Sol}) for $\mathcal{A}_n=\mathcal{F}$ where $\bar{\mathbf{c}}^{\mathcal{F}}=\bar{\mathbf{c}}$ and $\mathbf{s}_k^{\mathcal{F}}=\mathbf{s}_k$ hold. Note that this stationary point is the global optimal solution of the unconstrained version of the problem in (\ref{Eq:ML_Log}), i.e., when constraint $\bar{\mathbf{c}}\geq\mathbf{0}$ is dropped. Therefore, if $\bar{\mathbf{c}}^{\mathcal{F}}$ falls into the feasible set, i.e., $\bar{\mathbf{c}}^{\mathcal{F}}\geq\mathbf{0}$ holds, it is also the optimal solution of the constrained problem in (\ref{Eq:ML_Log}) and hence, we obtain $\hat{\bar{\mathbf{c}}}^{\mathtt{ML}}=\bar{\mathbf{c}}^{\mathcal{F}}$.

\textit{Boundary Point:} In this case, for the optimal solution, some of the elements of $\hat{\bar{\mathbf{c}}}$ are zero. Since it is not a priori known which elements are zero, we have to consider all possible cases. To do so, we use auxiliary variables $\bar{\mathbf{c}}^{\mathcal{A}_n}$ and $\mathbf{s}_k^{\mathcal{A}_n}$ where set $\mathcal{A}_n$ specifies the indices of the non-zero elements of $\bar{\mathbf{c}}$. For a given $\mathcal{A}_n$, we formulate a new problem by substituting $\bar{\mathbf{c}}^{\mathcal{A}_n}$ and $\mathbf{s}_k^{\mathcal{A}_n}$ for $\bar{\mathbf{c}}$ and $\mathbf{s}_k$ in (\ref{Eq:ML_Log}), respectively.  The stationary point of the new problem can be found by taking the derivative of  $g(\bar{\mathbf{c}}^{\mathcal{A}_n})$ with respect to $\bar{\mathbf{c}}^{\mathcal{A}_n}$ which is given in (\ref{Eq:ML_Sol}). Here, if $\bar{\mathbf{c}}^{\mathcal{A}_n}\geq \mathbf{0}$ does not hold, we discard $\bar{\mathbf{c}}^{\mathcal{A}_n}$, otherwise, it is a candidate for the optimal solution. Therefore, we  construct the candidate ML CIR estimate, denoted by $\hat{\bar{\mathbf{c}}}^{\mathtt{CAN}}$, such that the elements of $\hat{\bar{\mathbf{c}}}^{\mathtt{CAN}}$ whose indices are in $\mathcal{A}_n$ are equal to the values of the corresponding elements in $\bar{\mathbf{c}}^{\mathcal{A}_n}$ and the remaining elements are equal to zero. The resulting $\hat{\bar{\mathbf{c}}}^{\mathtt{CAN}}$ is saved in the candidate set $\mathcal{C}$. Finally, the ML estimate, $\hat{\bar{\mathbf{c}}}^{\mathtt{ML}}$, is given by that $\hat{\bar{\mathbf{c}}}^{\mathtt{CAN}}$ in set $\mathcal{C}$ which maximizes $g(\bar{\mathbf{c}})$.

The above results are concisely summarized in Algorithm~1 which concludes the proof.


\section{Proof of Lemma~\ref{Lem:LSSE}}\label{App:LSSE}

The optimization problem in (\ref{Eq:LSSE_Estimation}) is convex  since $\|\boldsymbol{\epsilon}\|^2$ is quadratic in variable $\bar{\mathbf{c}}$, $\mathbf{S}^T \mathbf{S} \succeq 0$ holds, and the feasible set $\bar{\mathbf{c}}\geq \mathbf{0}$ is linear in $\bar{\mathbf{c}}$ \cite[Chapter 4]{Boyd}. The constrained convex problem in (\ref{Eq:LSSE_Estimation}) can be solved using a similar methodology as was used for finding the ML estimate in Lemma~\ref{Lem:ML}. Hence, in order to avoid repetition and due to the space constraint, we provide only a sketch of the proof for Lemma~\ref{Lem:LSSE} in the following. In particular, the optimal LSSE estimate is either a stationary point or a boundary point for the problem in (\ref{Eq:LSSE_Estimation}). Moreover, if the solution is a boundary point for the original problem in (\ref{Eq:LSSE_Estimation}), we can formulate a new problem by substituting $\bar{\mathbf{c}}^{\mathcal{A}_n}$ and $\mathbf{s}_k^{\mathcal{A}_n}$ for $\bar{\mathbf{c}}$ and $\mathbf{s}_k$ in (\ref{Eq:LSSE_Estimation}), respectively, where the solution of the new problem has to be a stationary point, see the proof of Lemma~\ref{Lem:ML}.   For a given $\mathcal{A}_n$, the derivative of $\|\boldsymbol{\epsilon}^{\mathcal{A}_n}\|^2$, where $\boldsymbol{\epsilon}^{\mathcal{A}_n}=\mathbf{r}-\mathbf{S}^{\mathcal{A}_n} \bar{\mathbf{c}}^{\mathcal{A}_n}$, with respect to $\bar{\mathbf{c}}^{\mathcal{A}_n}$ is obtained as
\begin{IEEEeqnarray}{lll} \label{Eq:LSSE_Deriv}
 \frac{\partial \|\boldsymbol{\epsilon}^{\mathcal{A}_n}\|^2}{\partial \bar{\mathbf{c}}^{\mathcal{A}_n}}  = 2 (\mathbf{S}^{\mathcal{A}_n})^T \mathbf{S}^{\mathcal{A}_n}  \bar{\mathbf{c}}^{\mathcal{A}_n}  - 2  (\mathbf{S}^{\mathcal{A}_n})^T \mathbf{r} \overset{!}{=} 0,
\end{IEEEeqnarray}
where we employed the vector differentiation rules $\frac{\partial \mathrm{tr}\left\{ \mathbf{A}\mathbf{x}\mathbf{x}^T \right\}}{\partial \mathbf{x}} = \left(\mathbf{A}+\mathbf{A}^T\right)\mathbf{x}$ and $\frac{\partial \mathrm{tr}\left\{ \mathbf{A}\mathbf{x} \right\}}{\partial \mathbf{x}} = \mathbf{A}^T$ \cite[Table I]{TraceDerivative}. Solving (\ref{Eq:LSSE_Deriv}) leads to (\ref{Eq:LSSE_Sol}). Now, we first find the stationary point of the original problem by substituting  $\mathcal{A}_n=\mathcal{F}$ in (\ref{Eq:LSSE_Sol}). If $\bar{\mathbf{c}}^{\mathcal{F}}\geq \mathbf{0}$ holds, $\bar{\mathbf{c}}^{\mathcal{F}}$ is the optimal LSSE estimate, otherwise, we have to calculate  $\bar{\mathbf{c}}^{\mathcal{A}_n}$ for all possible boundary points, i.e., all elements of $\mathcal{A}_n$. Among all $\bar{\mathbf{c}}^{\mathcal{A}_n}$ for which $\bar{\mathbf{c}}^{\mathcal{A}_n}\geq \mathbf{0}$ holds, the LSSE estimate is the one  which minimizes $\|\boldsymbol{\epsilon}\|^2$. These results are summarized in Algorithm~1 which concludes the proof.

\section{Proof of Corollary~\ref{Corol:LSSE_Unbiased}}\label{App:LSSE_Consis}

To show that the LSSE estimator is biased in general, we neglect the constraint that  $\bar{\mathbf{c}} \geq \mathbf{0}$ has to hold in (\ref{Eq:LSSE_Estimation}) which yields an upper bound on the estimation error for the LSSE estimator, denoted by $\hat{\bar{\mathbf{c}}}^{\mathtt{LSSE}}_{\mathtt{up}}=\left(\mathbf{S}^T \mathbf{S}\right)^{-1}  \mathbf{S}^T \mathbf{r}$.    Thereby, we first show that $\hat{\bar{\mathbf{c}}}^{\mathtt{LSSE}}_{\mathtt{up}}$ is unbiased. The unbiasedness of $\hat{\bar{\mathbf{c}}}^{\mathtt{LSSE}}_{\mathtt{up}}$ is easily shown  as
\begin{IEEEeqnarray}{lll} \label{Eq:LSSE_Unbiased}
\mathbbmss{E}_{\mathbf{r}|\bar{\mathbf{c}}}\left\{ \hat{\bar{\mathbf{c}}}^{\mathtt{LSSE}}_{\mathtt{up}} \right \} 
& = \mathbbmss{E}_{\mathbf{r}|\bar{\mathbf{c}}}\left\{ \left(\mathbf{S}^T \mathbf{S}\right)^{-1}  \mathbf{S}^T \mathbf{r} \right \}
= \left(\mathbf{S}^T \mathbf{S}\right)^{-1}  \mathbf{S}^T \mathbbmss{E}_{\mathbf{r}|\bar{\mathbf{c}}}\left\{  \mathbf{r} \right \} \nonumber \\
& = \left(\mathbf{S}^T \mathbf{S}\right)^{-1}  \mathbf{S}^T  \mathbf{S}  \bar{\mathbf{c}} =  \bar{\mathbf{c}}. 
\end{IEEEeqnarray}
For the case when $\hat{\bar{\mathbf{c}}}^{\mathtt{LSSE}}_{\mathtt{up}}\geq\mathbf{0}$ holds,  the original $\hat{\bar{\mathbf{c}}}^{\mathtt{LSSE}}$ is identical to $\hat{\bar{\mathbf{c}}}^{\mathtt{LSSE}}_{\mathtt{up}}$. Otherwise, for the case when some of the elements of $\hat{\bar{\mathbf{c}}}^{\mathtt{LSSE}}_{\mathtt{up}}$ are negative,  $\hat{\bar{\mathbf{c}}}^{\mathtt{LSSE}}$ deviates from $\hat{\bar{\mathbf{c}}}^{\mathtt{LSSE}}_{\mathtt{up}}$ to ensure that $\hat{\bar{\mathbf{c}}}^{\mathtt{LSSE}}\geq\mathbf{0}$ holds. Therefore, the elements of $\mathbbmss{E}_{\mathbf{r}|\bar{\mathbf{c}}}\left\{ \hat{\bar{\mathbf{c}}}^{\mathtt{LSSE}} \right \} $ have a positive bias compared to their respective elements in $\mathbbmss{E}_{\mathbf{r}|\bar{\mathbf{c}}}\left\{ \hat{\bar{\mathbf{c}}}^{\mathtt{LSSE}}_{\mathtt{up}} \right \} $ 
 which indicates that the LSSE estimate, $\hat{\bar{\mathbf{c}}}^{\mathtt{LSSE}}$, is biased. 
 
To show that the LSSE estimator becomes asymptotically unbiased, as $K\to\infty$,  we first prove that $\hat{\bar{\mathbf{c}}}^{\mathtt{LSSE}}_{\mathtt{up}}$ is a consistent estimator, i.e., $\hat{\bar{\mathbf{c}}}^{\mathtt{LSSE}}_{\mathtt{up}}\to\bar{\mathbf{c}}$ as $K\to\infty$. Here, since $\hat{\bar{\mathbf{c}}}^{\mathtt{LSSE}}_{\mathtt{up}}$ provides an upper bound on the estimation error of $\hat{\bar{\mathbf{c}}}^{\mathtt{LSSE}}$, we can conclude that if $\hat{\bar{\mathbf{c}}}^{\mathtt{LSSE}}_{\mathtt{up}}$ is consistent, then $\hat{\bar{\mathbf{c}}}^{\mathtt{LSSE}}$ is also consistent. In other words, as $K\to\infty$, we obtain $\hat{\bar{\mathbf{c}}}^{\mathtt{LSSE}}\to\hat{\bar{\mathbf{c}}}^{\mathtt{LSSE}}_{\mathtt{up}}\to\bar{\mathbf{c}}$ and hence, the asymptotic unbiasness of the LSSE estimator. 

In the following, we employ a well-known approach which is based on the law of large numbers (LLN) to show the consistency  of $\hat{\bar{\mathbf{c}}}^{\mathtt{LSSE}}_{\mathtt{up}}$ \cite{Economy_OLS,IEEE_OLS}. To apply the LLN, we rewrite $\mathbf{S}=[\mathbf{l}_1,\mathbf{l}_2,\cdots,\mathbf{l}_L,\mathbf{l}_{\mathtt{n}}]$ where $\mathbf{l}_{l},\,\,l\in\{1,2,\dots,L,{\mathtt{n}}\}$, are the columns of $\mathbf{S}$. We note that for each element of vector $\mathbf{S}^T\mathbf{r} = [\mathbf{l}_1^T\mathbf{r},\mathbf{l}_2^T\mathbf{r},\cdots,\mathbf{l}_L^T\mathbf{r},\mathbf{l}_{\mathtt{n}}^T\mathbf{r}]$, we have a deterministic term $\mathbf{l}_{l},\,\,l\in\{1,2,\dots,L,{\mathtt{n}}\}$ and a stochastic term $\mathbf{r}$. Therefore, using the LLN, we obtain  $\frac{1}{K-L+1}\mathbf{l}_l^T\mathbf{r} \overset{K\to\infty}{\longrightarrow} \frac{1}{K-L+1} \mathbf{l}_l^T \mathbf{S}\bar{\mathbf{c}}$. Here, we emphasize that $\frac{1}{K-L+1} \mathbf{r} \overset{K\to\infty}{\longrightarrow}  \frac{1}{K-L+1} \mathbf{S}\bar{\mathbf{c}}$ does not hold. However, $\frac{1}{K-L+1} \mathbf{l}_l^T\mathbf{r}$ is indeed an averaging over all observations in which the element $\bar{c}_l$ is present based on (\ref{Eq:ChannelInOut}).  Substituting these results into  $\hat{\bar{\mathbf{c}}}^{\mathtt{LSSE}}_{\mathtt{up}}$, we obtain
\begin{IEEEeqnarray}{lll} \label{Eq:LLN}
\hat{\bar{\mathbf{c}}}^{\mathtt{LSSE}}_{\mathtt{up}}= \left(\mathbf{S}^T \mathbf{S}\right)^{-1}  \mathbf{S}^T \mathbf{r}
  \overset{K\to\infty}{\longrightarrow} \left(\mathbf{S}^T \mathbf{S}\right)^{-1}  \mathbf{S}^T \mathbf{S} \bar{\mathbf{c}} = \bar{\mathbf{c}}.
    \end{IEEEeqnarray}
Since $\hat{\bar{\mathbf{c}}}^{\mathtt{LSSE}}_{\mathtt{up}}$ is a consistent estimator and also provides an upper bound on the estimation error for   $\hat{\bar{\mathbf{c}}}^{\mathtt{LSSE}}$, we can conclude that  $\hat{\bar{\mathbf{c}}}^{\mathtt{LSSE}}$ is a consistent estimator, too. This completes the proof.

\section{Proof of Lemma~\ref{Lem:MMSE}}\label{App:MMSE}

The optimization problem in (\ref{Eq:MMSE_Estimation}) is a convex optimization problem since the cost function is quadratic in $\mathbf{F}$ \cite{Boyd}. Moreover, since the feasible set of $\mathbf{F}$ is not constrained, the global optimal $\mathbf{F}$ has to be a stationary point which can be found by equating the derivative of $\mathbbmss{E}_{\mathbf{r},\bar{\mathbf{c}}}\left\{\|\mathbf{e}^{\mathtt{MMSE}}_{\mathtt{up}}\|^2\right\}$ with respect to $\mathbf{F}$ to zero, i.e., 
\begin{IEEEeqnarray}{lll} \label{Eq:MMSE_Deriv}
 \frac{\partial \mathbbmss{E}_{\mathbf{r},\bar{\mathbf{c}}}\left\{\|\mathbf{e}^{\mathtt{MMSE}}_{\mathtt{up}}\|^2\right\}  }{\partial \mathbf{F}} \nonumber \\ 
 \quad =  \mathbf{F}\left(\boldsymbol{\Phi}_\mathbf{rr}+\boldsymbol{\Phi}_\mathbf{rr}^T\right)  
 -2 \boldsymbol{\Phi}_{\mathbf{r}\bar{\mathbf{c}}}^T  \nonumber \\
\quad = 2 \mathbf{F}\left(\mathbf{S}\boldsymbol{\Phi}_{\bar{\mathbf{c}}\bar{\mathbf{c}}} \mathbf{S}^T +  \mathrm{diag}\left\{\mathbf{S}\boldsymbol{\mu}_{\bar{\mathbf{c}}}\right\} \right) 
 -2\boldsymbol{\Phi}_{\bar{\mathbf{c}}\bar{\mathbf{c}}}^T\mathbf{S}^T   \overset{!}{=} \mathbf{0}. \quad
\end{IEEEeqnarray}
Solving the above linear equation leads to (\ref{Eq:MMSE_Sol}) and concludes the proof.

\bibliographystyle{IEEEtran}
\bibliography{Ref_12_08_2016}

\end{document}